\documentclass[sigconf]{acmart}

\AtBeginDocument{%
  }

\setcopyright{acmlicensed}
\copyrightyear{2025}
\acmYear{2025}
\setcopyright{cc}
\setcctype{by}
\acmConference[KDD '25]{Proceedings of the 31st ACM SIGKDD Conference on
Knowledge Discovery and Data Mining V.2}{August 3--7, 2025}{Toronto, ON, Canada}
\acmBooktitle{Proceedings of the 31st ACM SIGKDD Conference on Knowledge
Discovery and Data Mining V.2 (KDD '25), August 3--7, 2025, Toronto, ON,
Canada}
\acmDOI{10.1145/3711896.3736887}
\acmISBN{979-8-4007-1454-2/2025/08}

\usepackage{amsmath}
\usepackage{booktabs} 
\usepackage{multirow}
\usepackage{makecell}
\usepackage{xcolor}

\begin{document}

\title{CompilerDream: Learning a Compiler World Model for General Code Optimization}

\author{Chaoyi Deng}
\authornote{Both authors contributed equally to this research.}
\orcid{0009-0003-0635-2568}
\affiliation{%
  \department{School of Software, BNRist}
  \institution{Tsinghua University}
  \city{Beijing}
  \country{China}
}
\email{dengcy23@mails.tsinghua.edu.cn}

\author{Jialong Wu}
\authornotemark[1]
\orcid{0009-0008-7846-053X}
\affiliation{%
  \department{School of Software, BNRist}
  \institution{Tsinghua University}
  \city{Beijing}
  \country{China}
}
\email{wujialong0229@gmail.com}

\author{Ningya Feng}
\orcid{0009-0006-8448-2570}
\affiliation{%
  \department{School of Software, BNRist}
  \institution{Tsinghua University}
  \city{Beijing}
  \country{China}
}
\email{fny21@mails.tsinghua.edu.cn}

\author{Jianmin Wang}
\orcid{0000-0001-6841-7943}
\affiliation{%
  \department{School of Software, BNRist}
  \institution{Tsinghua University}
  \city{Beijing}
  \country{China}
}
\email{jimwang@tsinghua.edu.cn}

\author{Mingsheng Long}
\authornote{Corresponding author.}
\orcid{0000-0002-5412-9120}
\affiliation{%
  \department{School of Software, BNRist}
  \institution{Tsinghua University}
  \city{Beijing}
  \country{China}
}
\email{mingsheng@tsinghua.edu.cn}


\begin{abstract}
  Effective code optimization in compilers is crucial for computer and software engineering. The success of these optimizations primarily depends on the selection and ordering of the optimization passes applied to the code. While most compilers rely on fixed pass sequences, current methods to find the optimal sequence for specific programs either employ impractically slow search algorithms or learning methods that struggle to generalize to code unseen during training. To address these challenges, we introduce CompilerDream, the first world-model-based approach for general code optimization. CompilerDream features a compiler world model with a reward smoothing technique, enabling accurate simulation of optimization processes. Built on this model, code optimization agents can then be constructed via value prediction or direct optimization sequence generation. Trained on a large-scale program dataset, these agents serve as versatile code optimizers across diverse application scenarios and source-code languages. Our extensive experiments highlight CompilerDream's strong optimization capabilities for autotuning, where it leads the CompilerGym leaderboard. More importantly, the zero-shot generalization ability of large-scale trained compiler world model and agent, excels across diverse datasets, surpassing LLVM's built-in optimizations and state-of-the-art methods in both settings of value prediction and end-to-end code optimization.\footnote{Code and data available at \href{https://github.com/thuml/CompilerDream}{https://github.com/thuml/CompilerDream}.}
\end{abstract}




\maketitle

\ifdefempty{\kddavailabilityurl}{}{
\begingroup\small\noindent\raggedright\textbf{KDD Availability Link:}\\
The source code of this paper has been made publicly available at \url{https://doi.org/10.5281/zenodo.15552746}. The data of this paper has been made publicly available at \url{https://doi.org/10.5281/zenodo.15549673}.
\endgroup
}

\begin{figure}[h]
    \centering
    \includegraphics[width=1.0\linewidth]{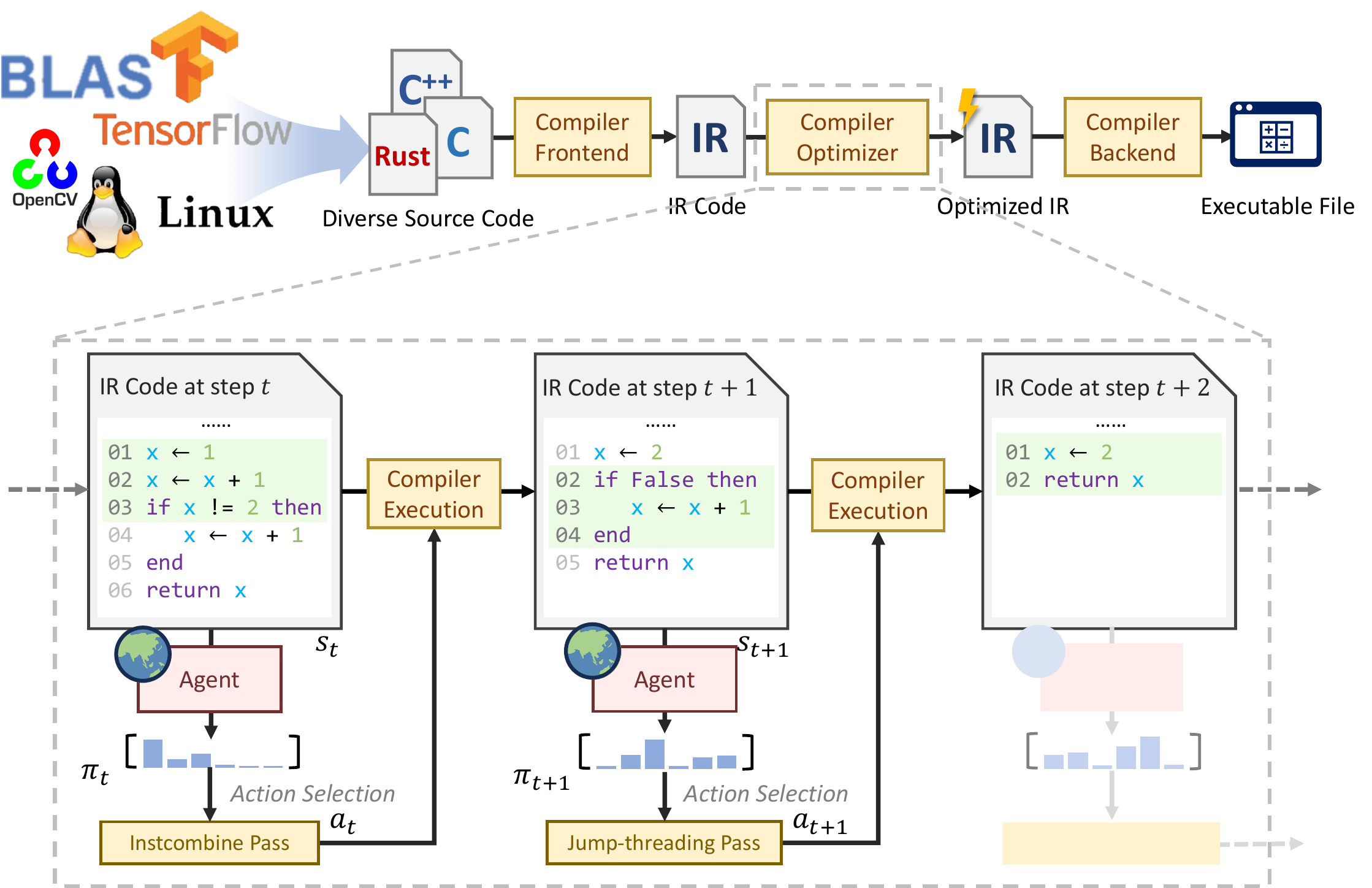}
    \caption{
    CompilerDream performs code optimization by interacting with the compiler using an agent powered by a world model. With strong generalization capabilities, it efficiently optimizes programs from diverse origins.
    }
    \Description{Diverse source code, including C++, C, Rust, and other languages, is first translated into an intermediate representation (IR) by the compiler frontend. CompilerDream, leveraging a world model, optimizes the IR by selecting an optimal pass at each step using an optimization agent. The optimized IR is then processed by the compiler backend to generate a high-quality executable file.}
    \label{fig:overview}
\end{figure}

\section{Introduction}

Code optimization plays an important role in realizing the full potential of software and hardware. Developers desire a universal solution to transform input programs into semantically equivalent but more efficient versions without manual effort. Compilers achieve this through a front-end that translates source code into an intermediate representation (IR), a middle-end optimizer that performs language- and platform-agnostic IR optimizations, and a back-end that converts IR to the binary code (Figure \ref{fig:overview}). The optimizer is typically implemented as a series of \textit{passes} applying transforms on the code, where performance largely depends on the selection and order of these optimization passes. Standard compilers use a few fixed sets of optimization sequences to enhance specific aspects of program performance, such as \texttt{-O1}, \texttt{-O2}, and \texttt{-O3} for execution speed, and \texttt{-Os} and \texttt{-Oz} for program size reduction.

Obviously, given the vast diversity of programs and platforms, these off-the-shelf strategies predefined by compiler experts are suboptimal for most circumstances. Automatically optimizing the pass sequence for specific programs thus can yield significant performance gains over default compiler settings \cite{triantafyllis2003compiler, georgiou2018less}. To be practical, such an algorithm must produce a satisfactory pass sequence within a reasonable time and handle a wide variety of programs. However, current research often fails to meet these requirements simultaneously. Search-based methods \cite{bodin1998iterative} achieve near-optimality but require thousands of compilations per program to validate the optimization outcomes, making them impractical. In contrast, machine learning methods avoid these time-consuming compiler interactions, by either predicting the optimization sequences directly or estimating the outcomes of optimization sequences to guide a search. 

However, these learning-based methods still risk sacrificing some optimality and, more importantly, face a significant bottleneck in broad generalization across diverse programs that may be out of the training samples. A range of machine learning applications \cite{devlin2018bert, brown2020language, radford2021learning, kirillov2023segment} have witnessed that training high-capacity models on large-scale datasets can yield unprecedented performance. Prior studies on compiler optimization also suggest that training on large datasets with diverse programs could be beneficial \cite{cummins2022compilergym}, yet the prevalent practice in this field is still to learn optimization strategies in a per-program manner \cite{huang2019autophase, shahzad2022reinforcement} or from relatively small training sets comprising only a few hundred programs with limited-capacity models \cite{mammadli2020static, jain2022poset}, which hinders generalization.

This paper focuses on the LLVM \cite{lattner2004llvm} phase ordering problem, a longstanding challenge for compiler research. We propose CompilerDream, a world-model-based approach for general code optimization, capable of handling a wide variety of programs, unlike most prior approaches focusing on narrow, domain-specific ones. It learns an accurate predictive world model to simulate compiler executions, forecasting future IR states and optimization metric improvements based on the current IR state and applied pass. We believe that employing a world model offers the following advantages for compiler optimization: First, by capturing optimization dynamics, the world model gains generalizable knowledge into our method \cite{anand2021procedural}, improving the overall generalization performance. Second, it replaces costly compiler invocations, significantly reducing computational overheads of search and learning algorithms and facilitating large-scale training. Additionally, its high-capacity architecture can extract deeper insights from extensive datasets.

Specifically, to better adapt the world model for compiler optimization, we closely examine the optimization process and introduce a reward smoothing technique to enhance world model training and improve simulation accuracy. Built on this accurate world model, CompilerDream supports various types of optimization agents, including value prediction and reinforcement learning. Moreover, we carefully curate a large-scale training dataset of natural programs, enhancing the world model's and agent's generalization to unseen programs, enabling our method to predict superior optimization sequences.

We demonstrate CompilerDream's effectiveness across a range of program domains \cite{cummins2022compilergym} and problem scenarios. Our test domains include benchmark suites covering fundamental algorithms, as well as production-level open-source programs, such as object files from C++ TensorFlow \cite{abadi2016tensorflow} and OpenCV \cite{culjak2012brief} libraries. Even without considering generalization, CompilerDream's strong optimization capabilities top the CompilerGym leaderboard for autotuning. For general code optimization, our large-scale trained world model accurately predicts the outcomes of pass sequences on unseen programs. Both types of optimization agents equipped with the world model outperform the built-in \texttt{-Oz} flag and state-of-the-art methods in their respective settings. In the value prediction setting, the world model serves as an accurate value predictor, enabling superior action sequence selection. In the reinforcement learning setting, the agent trained entirely in the world model directly generates optimization sequences in a single trial, achieving more efficient code across diverse datasets. 

The main contributions of this work are as follows:
\begin{itemize}
    \item We propose \textit{CompilerDream}, the first world-model-based approach for code optimization.
    \item We introduce a \textit{reward smoothing technique} that facilitates the application of world models to compiler optimization, benefiting future research in this area.
    \item Our approach supports \textit{multiple optimization agents}, including value prediction and reinforcement learning agents, unifying different methods in this field.
    \item By leveraging the \textit{large-scale CodeContests dataset} and the world model's generalization capability, we achieve strong performance on diverse unseen programs.
    \item Extensive experiments across various program domains and problem scenarios, including autotuning, value prediction, and end-to-end code optimization, show that CompilerDream outperforms state-of-the-art methods.
\end{itemize}

\section{Related Work}
\label{related_work}

A key challenge in compilation is determining which code transformations to apply, how to apply them (e.g., using suitable parameters), and in what order.  This involves effectively searching and evaluating numerous options, a process known as iterative compilation \cite{bodin1998iterative} or autotuning \cite{datta2008stencil}. However, this search-based approach only finds a good optimization for one specific program and does not generalize into a compiler strategy. This limitation underscores the importance of integrating machine learning techniques.

\paragraph{\textbf{Supervised learning.}} Pioneering work has delved into supervised machine learning, adopting two main approaches \cite{leather2020machine}. The first approach requires an extensive search for each training program to identify the most effective optimization sequence, which then serves as the data labels. An early example \cite{calder1997evidence} used a neural network for branch prediction, and one more well-known work is MilepostGCC \cite{fursin2008milepost}, a practical attempt to integrate machine learning into a production compiler, GCC. It employs models trained on a large dataset of programs distributed over the Internet. The second approach aims to learn a cost or performance function capable of estimating the quality of various compiler options, which enables the evaluation of possible options without the need to compile and profile each one \cite{stephenson2003meta, luk2009qilin}. Coreset-NVP \cite{liang2023rlcompopt} follows this approach and achieves state-of-the-art performance.

\paragraph{\textbf{Reinforcement learning.}} Recent advancements have seen reinforcement learning (RL) techniques making strides in compiler optimization, circumventing the need for collecting optimal labeled data \cite{kulkarni2012mitigating}. This technique has been applied to optimize individual compilation heuristics, such as inlining \cite{trofin2021mlgo}, loop transformation \cite{haj2020neurovectorizer, brauckmann2021reinforcement}, and graph partitioning \cite{mirhoseini2017device}. Several works relevant to us, including AutoPhase \cite{huang2019autophase}, CORL \cite{mammadli2020static}, and POSET-RL \cite{jain2022poset}, have explored the full optimization pipeline, i.e., the phase ordering problem, using model-free RL without incorporating a world model.

\paragraph{\textbf{World model.}} A world model \cite{ha2018recurrent, lecun2022path} that approximates state transitions and reward signals is typically used in two ways: (1) it enables \textit{simulation} of "unseen" interactions that are unavailable or unaffordable \cite{janner2019trust, hafner2019dream}; (2) as an auxiliary learning task, it aids in better \textit{representation} that captures the underlying structure of the environment \cite{anand2021procedural, mazoure2021cross, ma2024harmonydream}. To the best of our knowledge, we are the first to introduce world models for code optimization. In this context, a world model can function as a compiler simulator, approximating IR transformations and eliminating the need for costly execution and profiling of extensive optimization sequences. Furthermore, by sharing representations with the world model, the policy can generalize more effectively to unseen programs. Thus, model-based agents \cite{sutton2018reinforcement} can be implemented, which offer superior sample efficiency and generalization compared to model-free methods.

\section{Method}

This section first defines the code optimization problem and design choices for observation, action, and reward (Section~\ref{sec:pomdp}). We then present the design and training technique of CompilerDream's world model (Section~\ref{sec:model-based_method}), two optimization agents (Section~\ref{sec:applications}), and considerations for large-scale dataset curation (Section~\ref{sec:train_dataset}).

\subsection{Phase Ordering Decision Process}

\label{sec:pomdp}
As illustrated in Figure~\ref{fig:overview}, one key problem of compiler optimization is to find the optimal sequence of optimization passes for a given program, also known as the \textit{phase ordering} problem. It can be naturally formulated as a \textit{partially observable Markov decision process} (POMDP) $\mathcal{M} = (\mathcal{S}, \mathcal{A}, r, p, \mu, \mathcal{O}, \phi)$ \cite{sutton2018reinforcement}. The state space $\mathcal{S}$ covers all possible Intermediate Representations (IRs), the action space $\mathcal{A}$ comprises individual compiler optimization passes, and the reward function $r$ is defined by the metric being optimized. The transition dynamics $p: \mathcal{S}\times \mathcal{A} \mapsto \mathcal{S}$ represents the outcome of applied IR transformations. The initial state distribution $\mu \in \Delta(\mathcal{S})$ captures all IRs of interest, which can be approximated via uniform sampling from the training dataset. The observation function $\phi: \mathcal{S} \mapsto \mathcal{O}$ maps the underlying IR into the observation space, capturing useful features.

A \textit{code optimization agent} determines the optimization sequence for an input program through interactions with the compiler environment. At each time step $t=0,1,2,\dots$, the agent applies an action $a_t$ to current IR based on observation $o_t$, transforming current state to $s_{t+1} = p(s_t, a_t)$ and receiving reward $r_t$. The agent can employ effective strategies, such as search algorithms and parametric neural networks, to select the best actions, as detailed in Section~\ref{sec:applications}.

Under the POMDP formulation, we design the \textit{observation} and \textit{action space} as summarized in Table~\ref{tab:obs_action}. The observation features were selected for efficiency and effectiveness. 
In preliminary experiments, we find that complex program features like ProGraML \cite{cummins2020programl} and inst2vec \cite{ben2018neural} significantly slow down CompilerDream with marginal performance gains. In contrast, expert-designed Autophase features \cite{huang2019autophase} leverage domain knowledge, enhancing generalization by filtering irrelevant details. We construct the observation by concatenating the 56-dimensional Autophase feature vector with a 42-dimensional action histogram vector, which records the number of times each action has been selected during the current episode. Both vectors are normalized for consistency: each Autophase feature is divided by the program’s initial total instruction count, and the action histogram is scaled by the per-episode action limit, which is set to 45.
We adopt two distinct action spaces in our experiments to align with the baseline methods. The full action space consists of all 124 LLVM optimization passes, while the reduced action space derived from Autophase \cite{huang2019autophase} consists of 42 actions. Originally, this action space included 45 LLVM passes, but CompilerGym excludes 3 due to updates in the latest LLVM version, leaving 42 actions.  This reduced action space is widely used and has been shown to be effective in prior studies \cite{huang2019autophase, cummins2022compilergym}. 

The \textit{reward function} is defined as the normalized change of the optimization metric $C(s)$:
\begin{equation}
    \label{f:metric}
    r_{t+1} = \frac{C(s_{t}) - C(s_{t+1})}{C(s_0) -C(s_\mathrm{b})},
\end{equation}
where lower $C$ indicates better performance. Following prior work on code size reduction, we define $C(s)$ as the IR instruction count. $C(s_\mathrm{b})$ is the baseline performance achieved by the built-in \texttt{-Oz} flag.

\begin{table}[t]
\centering
\caption{Observation and action space in CompilerDream.}
\label{tab:obs_action}
\setlength\tabcolsep{2pt}
\begin{tabular}{lclc}
\toprule
\textbf{Aspect}                                                                                   & \multicolumn{1}{l}{\textbf{Name}}                          & \textbf{Description}                                                                                                                                                                      & \textbf{Dim.} \\ \midrule
\makecell[l]{\textbf{Observation}}         & \begin{tabular}[c] {@{}c@{}} Autophase\\ \cite{huang2019autophase}\end{tabular}       & \begin{tabular}[c]{@{}l@{}}A feature vector capturing \\ various IR code statistics.\end{tabular}                                                                                                                                    & 56                 \\ \cmidrule{2-4} 
  & \begin{tabular}[c]{@{}c@{}}Action\\ Histogram\end{tabular} & \begin{tabular}[c]{@{}l@{}}A vector where each \\ dimension represents  the \\ execution frequency of a \\ specific action.\end{tabular} & \begin{tabular}[c] {@{}c@{}} 124 \\ /42\end{tabular}          \\ \midrule
\begin{tabular}[c]{@{}l@{}}\textbf{Action}\\ (Sec. 4.2 \& 4.3)\end{tabular} & \begin{tabular}[c]{@{}c@{}}Full LLVM\\Passes\end{tabular} & \begin{tabular}[c] {@{}l@{}} Full action space with all \\ LLVM optimization passes. \end{tabular}                                                                                                                                     & 124                \\\cmidrule{2-4} 
\begin{tabular}[c]{@{}l@{}}\textbf{Action}\\ (Sec. 4.4)\end{tabular}        & \begin{tabular}[c]{@{}c@{}}Autophase\\ Passes\end{tabular} & \begin{tabular}[c]{@{}l@{}}A reduced action space \\ derived from Autophase\cite{huang2019autophase}.\end{tabular}                                                                                                  & 42                 \\ \bottomrule
\end{tabular}
\end{table}

\begin{figure*}[tbp]
    \centering
    \includegraphics[width=1.0\linewidth]{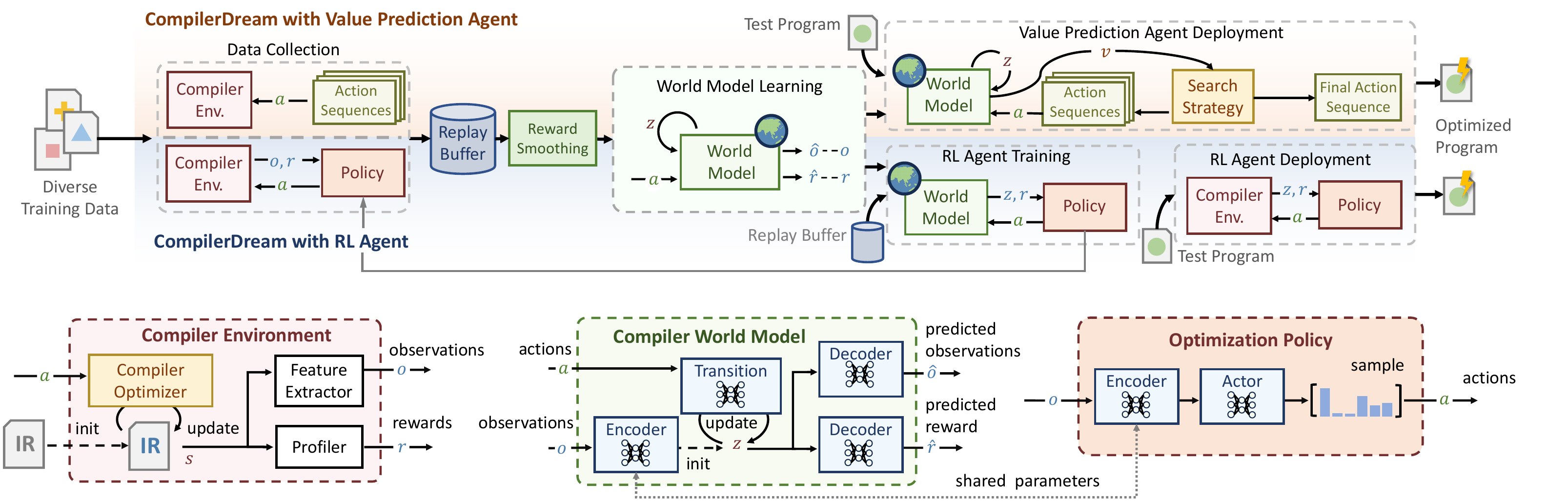}
    \caption{
    Design overview of CompilerDream. The world model is trained on real episodes generated by the agent, incorporating a reward smoothing process to stabilize training and enhance learning efficiency. CompilerDream supports training with RL and value prediction agents, which learn from episodes simulated by the world model. 
    }
    \Description{The agent interacts with a compiler to collect real episodes, which are stored in a replay buffer. While training the world model with these episodes, a reward smoothing process is applied to enhance training stability and learning efficiency. Once trained, the world model functions as a compiler environment simulator, enabling reinforcement learning (RL) agents and value prediction agents to interact with it and generate simulated episodes. These simulated episodes are used to train the agents, ultimately leading to improved performance.}
    \label{fig:method}
\end{figure*}

\subsection{Learning a Compiler World Model}
\label{sec:model-based_method}

Following the advanced Dreamer approach \cite{hafner2023mastering}, we build a world model to learn the formulated POMDP process of compiler optimization, as depicted in Figure \ref{fig:method}. Concretely, we train a model $(\hat p_\theta, \hat r_\theta )$ of the compiler environment parameterized by $\theta$, which approximates the underlying transition dynamics of optimization process $ p(o_{t+1} | o_{\leq t}, a_{\leq t})$  and the reward function $r(o_{\leq t}, a_{\leq t})$.

The compiler world model simulating the compiler environment is formulated with the following four components: 
\begin{equation}
\begin{aligned}
&\text{Representation model:} &&z_t\sim q_\theta(z_{t} \mid z_{t-1},a_{t-1}, o_{t}), \\
&\text{Transition model:} &&\hat{z}_t\sim p_\theta(\hat{z}_{t} \mid z_{t-1}, a_{t-1}), \\
& \text{Observation decoder:} &&\hat{o}_t\sim p_\theta(\hat{o}_{t} \mid z_{t}),\\
& \text{Reward decoder:} &&\hat{r}_t \sim p_\theta(\hat{r}_t  \mid  z_t).
\label{eq:dreamer}
\end{aligned}
\end{equation}

The representation model estimates a \textit{neural compiler state} $z_t$ from the current observation $o_t$ of \textit{real compiler state}, the previous state $z_{t-1}$ and the previous optimization action $a_{t-1}$. A representation loss $\mathcal{L}_\mathrm{repr}$ is used to train the neural compiler states to accurately reconstruct the observation and reward by two decoders:
\begin{align}
\label{f:loss_cons}
    &\mathcal{L}_{\mathrm{repr}}(\theta) \doteq -\ln p_{\theta}(o_{t}| z_{t})  -\ln p_{\theta}(r_{t}| z_{t}).
\end{align}

The transition model captures the dynamics of the compiler world model, predicting future neural compiler state ${\hat z}_t$ directly from $z_{t-1}$ and $a_{t-1}$. 
A prediction loss minimizes the difference between the estimated neural compiler state $z_t$ and the predicted compiler state ${\hat z}_t$, simultaneously enhancing the transition model's accuracy in predicting future states and making the neural compiler state produced by the representation model easier to predict:
\begin{align}
\label{f:loss_pred}
    &\mathcal{L}_{\mathrm{pred}}(\theta) \doteq \mathrm{KL}\left[q_{\theta}(z_{t}| z_{t-1},a_{t-1}, o_{t}) \,\Vert\,p_{\theta}(\hat{z}_{t}| z_{t-1}, a_{t-1}) \right].
\end{align}

The overall models are jointly learned by minimizing the sum of the representation loss and the prediction loss.

\paragraph{\textbf{Compiler simulation}} We can simulate the behavior of the real compiler using our trained compiler world model. Specifically, a \emph{simulated} compiler optimization trajectory $\{ \hat{z}_\tau, \hat{a}_\tau, \hat{r}_\tau, \hat{o}_\tau\}$ with horizon $H$ can be generated by the interactions between the world model and an optimization agent: starting at a neural state $\hat{z}_t \sim q_\theta(z_t| o_t)$, at each step $\tau=t, t+1, t+2, \dots$, the agent takes an action $\hat{a}_\tau \sim \pi_\psi \left(\hat{a}_\tau| \hat{z}_{\tau}\right)$, and transits to the next latent state $\hat{z}_{\tau+1}\sim p_\theta(\hat{z}_{\tau+1} | z_{\tau}, a_{\tau})$ with a reward $\hat{r}_{\tau+1} \sim p_\theta(\hat{r}_{\tau+1} |  \hat{z}_{\tau+1})$. Predicted real compiler state $\hat{o}_{\tau+1}$ can be optionally reconstructed by the observation decoder.

\begin{figure}[b]
    \centering
    \includegraphics[width=0.9\linewidth]{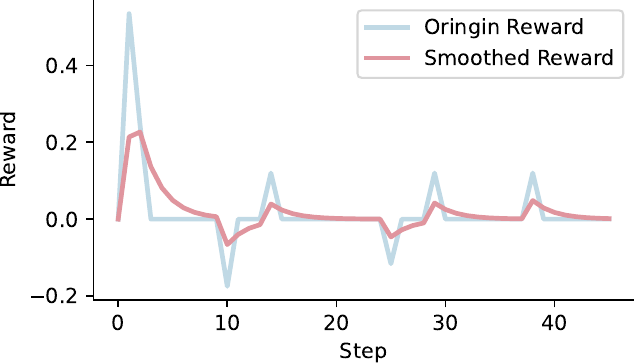}
    \caption{Comparison of the original and smoothed rewards during an episode on a program from the BLAS dataset \cite{lawson1979basic}.}
    \Description{Taking a typical episode as an example: out of 45 actions, only 8 receive non-zero rewards before smoothing, and they are scattered throughout the optimization process. After smoothing, most rewards become non-zero, and abrupt changes between adjacent rewards are eliminated. Moreover, the overall trend remains consistent with the original reward distribution.}
    \label{fig:smooth}
\end{figure}

\paragraph{\textbf{Reward smoothing}}
We discover that most optimization passes in an optimization sequence do not change the program IR instruction count, leading to \textit{sparse} rewards. Despite this, these passes are essential as they may modify critical properties, such as instruction order or replacements, that affect the effectiveness of subsequent passes. Furthermore, improvements tend to \textit{saturate} over time, as large non-zero rewards typically appear early in the sequence and diminish as optimization progresses.

In Figure~\ref{fig:smooth}, we present a typical episode of the optimization process: out of 45 actions, only 8 yielded non-zero rewards, whose values decrease as optimization progresses. These properties pose challenges for training the agent effectively, as it receives little to no guidance from most actions, even when some are vital to the final result. Moreover, the world model may struggle to predict reward values accurately, often defaulting to zero. This challenge arises from the significant class imbalance between zero and non-zero rewards, making it difficult to determine both the timing and magnitude of non-zero rewards.

Therefore, we applied reward smoothing to mitigate their sparsity and long-tailed distribution within an episode. This is achieved by adding an exponential decay to rewards: 
\begin{equation}
    r_t^\prime \leftarrow \alpha r_{t-1}^\prime + (1-\alpha) r_t,\;\; t=1,2,\dots
\end{equation}
with $\alpha \in [0,1)$. Consequently, we train a reward decoder $p_\theta(\hat{r}_t^\prime\mid  z_t)$ to predict the smoothed rewards. As shown in Figure~\ref{fig:smooth}, the smoothed rewards provide non-zero feedback for most steps, making them easier for CompilerDream to learn. It is worth noting that the total reward of an episode remains approximately unchanged after smoothing. For all \( t \geq 1 \), its contribution to the total reward after smoothing approximates the original reward value if the horizon is infinite: $\lim_{n\to\infty}(1-\alpha)\sum_{i=1}^n\alpha^i r_t=r_t$. Since most episodes exhibit a long-tailed reward distribution and we use a relatively small smoothing factor (\( \alpha = 0.6 \)), the smoothed total reward is expected to closely match the original total reward.

\subsection{Learning a Code Optimization Agent}
\label{sec:applications}

Building on the world model, we implement two agent designs, unifying supervised and reinforcement learning methods for compiler optimization into our world-model-based approaches.

\paragraph{\textbf{Value prediction agent}}
This kind of agent adopts a classic approach in compiler optimization, employing a heuristic to guide search methods \cite{liang2023rlcompopt, luk2009qilin, stephenson2003meta}. It comprises a \textit{value prediction model} $v_{\theta}$ to estimate the effect of an optimization sequence and a \textit{search strategy} that leverages this model to identify the best sequence. 

The \textit{value} of a program state $s$ is defined as the expected cumulative reward of applying an action sequence to the program. This aligns with the ratio of the reduced IR instruction count achieved between applying action sequence $\{a_\tau\}_{\tau=1}^m$ and standard $-Oz$ flag, according to the reward function $r$ defined in Eq.~\eqref{f:metric}:
\begin{equation}
    \label{f:value}
    v(s_0, \{a_\tau\}) = \sum_{\tau=1}^{m} r_\tau = \sum_{\tau=1}^{m}\frac{C(s_{\tau-1})-C(s_{\tau})}{C(s_0)-C(s_b)}=\frac{C(s_{0})-C(s_{m})}{C(s_0)-C(s_b)}.
\end{equation}

We leverage our world model to build a \textit{value prediction model}. As shown Section~\ref{sec:model-based_method}, from an initial observation $o_0$, and an action sequence $\{a_\tau\}$, our world model can simulate an optimization trajectory \( \{ \hat{z}_\tau, \hat{a}_\tau, \hat{r}_\tau, \hat{o}_\tau \}_{\tau=1}^m \). Summing up all simulated rewards $\hat{r}_\tau$ yields the predicted value: 
\begin{equation}
    v_{\theta}(s_0, \{a_\tau\})=\sum_{\tau=1}^m\hat{r}_\tau\approx\sum_{\tau=1}^mr_\tau
\end{equation} 

The search strategy then seeks the optimal optimization sequence $\{a_\tau\}^*$ by evaluating candidates using the value prediction model $v_\theta$. Although various search algorithms are applicable, we employ a simple strategy that enumerates action sequences within a fixed search space (detailed in Section~\ref{sec:coreset_exp}).

\paragraph{\textbf{Reinforcement learning agent}}
RL is another popular approach to compiler optimization, learning a policy to select passes sequentially. Unlike prior methods that rely on costly compiler interactions, CompilerDream trains RL agents on world model-simulated compiler optimization trajectories, significantly improving efficiency.

Our RL agent comprises an actor and a critic neural networks, both parameterized based on the neural compiler state:
\begin{align}
\begin{split}
    &\text { Actor: } \hat{a}_t \sim \pi_\psi\left(\hat{a}_t \mid \hat{z}_t\right) \\ &\text { Critic: } v_{\xi}\left(\hat{z}_t\right) \approx \mathbb{E}_{p_\theta, \pi_\psi}\left[\sum\nolimits_{\tau \geq t} \gamma^{\tau-t} \hat{r}_\tau\right].
\end{split}
\end{align}

\begin{figure*}[tbp]
    \centering
    \includegraphics[width=0.99\linewidth]{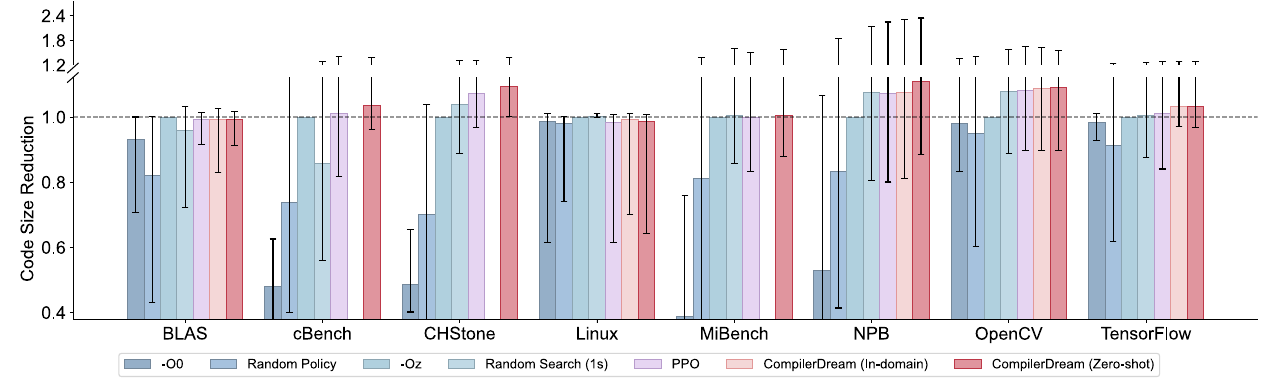}
    \vspace{5pt}
    \caption{Results of general code optimization with RL: Code size reduction in terms of IR instruction count over LLVM \texttt{-Oz} under different methods. Bars indicate the geometric mean and min-max range across test programs in each benchmark dataset.}
    \Description{The performance of several methods (-O0, Random Policy, -Oz, Random Search, PPO, In-domain CompilerDream, and Zero-shot CompilerDream) on different benchmarks (BLAS, cBench, CHStone, Linux, MiBench, NPB, OpenCV and TensorFlow).}
    \label{fig:results}
\end{figure*}

The critic evaluates the $\gamma$-discounted value $v_\xi(\hat{z}_t)$ of simulated neural state $\hat{z}_t$ under policy $\pi_{\psi}$. It is trained by minimizing the difference between the predicted value $v_\xi(\hat{z}_t)$ and the Monte-Carlo or more advanced bootstrapped return \cite{sutton2018reinforcement}, denoted as $V_t$:
\begin{align}
    &\mathcal{L}_{\text{critic}}(\xi)\doteq\mathbb{E}_{p_{\theta}, \pi_\psi}\left[\sum^{t+H}_{\tau=t} - \log v_{\xi}(V_{\tau}\mid \hat{z}_{\tau}) \right], 
    \label{eq:critic_loss}
\end{align}

The actor produces an optimization policy $\pi_\psi$ that predicts an action distribution of the best pass to choose, which is trained to maximize the simulated return through the REINFORCE policy gradient \cite{williams1992simple} with an entropy regularization \cite{haarnoja2018soft}:
\begin{align}
    \nonumber\mathcal{L}_{\text{actor}}(\psi)\doteq \mathbb{E}_{p_{\theta}, \pi_\psi} \Big[\sum^{t+H}_{\tau=t} \Big(&-\left(V_\tau -v_\xi(\hat{z}_{\tau})\right)\log \pi_\psi (\hat{a}_\tau\mid \hat{z}_{\tau}) \\&- \eta\,\mathbb{H}\left[\pi_\psi (\hat{z}_{\tau}) \right] \Big) \Big],
\end{align}

\subsection{Data Curation}
\label{sec:train_dataset}

To facilitate that our \textit{CompilerDream} method can effectively generalize to unseen situations, a concept known as zero-shot generalization, we have identified three critical factors in preparing our training dataset. First, the dataset should reflect \textit{naturalness}. During our preliminary experiments, we found that those generated code datasets like Csmith \cite{yang2011finding} and llvm-stress \cite{lattner2004llvm} provide no benefits or even hurt the generalization to real-world scenarios. Second, the dataset should be \textit{large} to prevent the agent from overfitting to a small number of codes and failing to generalize. Last, the code must exhibit \textit{high quality} from an optimization perspective. We need data with complex algorithmic logic and potential for optimizations to help the world model and the agent better understand the problem's intricacies. Datasets like AnghaBench \cite{da2021anghabench} which consist of millions of human-written codes collected from GitHub, are often too simple to allow for significant optimization improvements.

\begin{table}[tb]
\centering
\caption{Comparison of different training datasets for compiler optimization.}
\label{tab:datasets_compare}
\setlength{\tabcolsep}{1.5mm}
\begin{tabular}{lllll}
\toprule
\begin{tabular}[l]{@{}l@{}}\textbf{Dataset}\\ \textbf{Name}\end{tabular} &
  \begin{tabular}[l]{@{}l@{}}\textbf{Number of} \\ \textbf{IR files}\end{tabular} &
  \begin{tabular}[l]{@{}l@{}}\textbf{Large-} \\ \textbf{scale}\end{tabular} &
  \begin{tabular}[l]{@{}l@{}}\textbf{Human-} \\ \textbf{written}\end{tabular} &
  \begin{tabular}[l]{@{}l@{}}\textbf{Code} \\ \textbf{quality}\end{tabular} \\
  \midrule
cBench \cite{fursin2007midatasets}                & 23                 & No           & Yes          & High          \\
Mibench \cite{guthaus2001mibench}               & 40                 & No           & Yes          & Low           \\
Csmith \cite{yang2011finding}               & $\infty$                               & Yes          & No           & High          \\
llvm-stress \cite{lattner2004llvm}          & $\infty$                               & Yes          & No           & Low           \\
AnghaBench \cite{da2021anghabench}           & $1,041,333$                            & Yes          & Yes          & Low           \\ 
\textbf{CodeContests} \cite{li2022competition} & \textbf{$110,240$} & \textbf{Yes} & \textbf{Yes} & \textbf{High}\\\bottomrule
\end{tabular}
\end{table}

Therefore, we choose to construct our training datasets on top of the CodeContests dataset released with AlphaCode \cite{li2022competition}, which consists of over 13,000 problems of coding competition, and each problem, on average, has hundreds of solutions in multiple languages. We subsample up to ten C++ solutions for each training problem, resulting in 110,240 programs, as our training data, and sample one solution for each of 100 test problems as our validation data. As shown in Table~\ref{tab:datasets_compare}, CodeContests meets all our criteria, whereas other datasets fall short in one or more aspects.

\section{Experiments}
\label{sec:exp}

We conducted extensive experiments across various settings, comparing CompilerDream against state-of-the-art methods. Our study aims to answer the following key research questions (RQs):

\begin{enumerate}
    \renewcommand{\labelenumi}{\textbf{RQ\ \theenumi}\ }
    \item \textit{Optimality}: \label{rq:opt}Can CompilerDream's joint training process of the compiler world model and optimization agent discover superior optimization sequences?
    \item \textit{Accuracy}: \label{rq:acc}Does CompilerDream's world model produce accurate simulations of the real optimization process?
    \item \textit{Generalization}: \label{rq:gen}Is the policy learned with the world model still effective on various unseen programs?
    \item \textit{Effectiveness}: \label{rq:abl}Are all the techniques employed in CompilerDream effective in enhancing optimization results?
\end{enumerate}

In the following sections, we first show that CompilerDream excels as a powerful autotuning method, leading the CompilerGym leaderboard by discovering superior optimization sequences and leveraging accurate world model (Section \ref{sec:search_exp}), addressing RQ~\ref{rq:opt} and partially RQ~\ref{rq:acc}. We then demonstrate that its value prediction agent (Section \ref{sec:coreset_exp}) and RL agent (Section \ref{sec:main_exp}) both outperform state-of-the-art methods in various unseen test datasets, addressing RQ~\ref{rq:acc} and RQ~\ref{rq:gen}. Finally, Section~\ref{sec:analysis} presents ablation studies for RQ~\ref{rq:abl}.

\subsection{Evaluation}
\label{sec:eval}
Our experiments focus on code size reduction, benefiting applications on low-resource hardware like embedded systems. This focus stems from the practical advantages of code size as a metric: It is cost-effective to construct compilable training and test datasets and to evaluate the optimization performance for code size.

\paragraph{\textbf{Metric}}
To be more robust to outliners, we evaluate the code size optimization results by the geometric mean of IR instruction count reduction of program $s$ in a dataset $\mathcal{D}$:
\begin{equation}
\label{f:metric_geom}
R_{\mathcal{D}}(\text{agent}) = \left(\prod_{s\in\mathcal{D}}\frac{C(s_{b})}{C(s_\text{agent})}\right)^{\frac{1}{|\mathcal{D}|}}
\end{equation}
where $C$ denotes the IR instruction count, $s_{b}$ is the IR state optimized by LLVM's \texttt{-Oz} flag serving as a baseline, and $s_\text{agent}$ is the final IR state produced by agent. A value of $R$ above $1$ indicates superior performance compared to LLVM's \texttt{-Oz} option.

\paragraph{\textbf{Benchmarks}}
We evaluate our method mainly on benchmarks from the CompilerGym platform \cite{cummins2022compilergym}: benchmark suites including cBench \cite{fursin2007midatasets}, CHStone \cite{hara2008chstone}, MiBench \cite{guthaus2001mibench}, and NASA Parallel Benchmarks (NPB) \cite{bailey1995parallel}, as well as kernels from open source projects such as BLAS \cite{lawson1979basic}, Linux, OpenCV \cite{culjak2012brief}, and TensorFlow \cite{abadi2016tensorflow}.  Synthetic benchmarks from program generators \cite{lattner2004llvm, yang2011finding} are excluded as they lack real-world relevance. We adhere to the standard data splits of CompilerGym. For benchmarks with a total number of programs more than 100, we use the first 50 programs as the test set, the following 50 programs as the validation set, and all of the rest as the training set. These training and validation sets are only used for in-domain training. The datasets comprising fewer than 100 programs are not applicable for in-domain training; instead, all their programs are allocated to the test set. The number of programs in each dataset after division is detailed in Table \ref{tab:compilergym-dataset}. 

\begin{table}[t]

\centering
\caption{Dataset division of 8 CompilerGym benchmarks.}
\label{tab:compilergym-dataset}
\setlength{\tabcolsep}{1mm}
\begin{tabular}{lcccc}
\toprule
\textbf{Benchmark}    &\textbf{Training Split} & \textbf{Validation Split}    & \textbf{Test Split} \\
\midrule
BLAS       & 200           & 50                  & 50            \\
cBench     & N/A           & N/A                 & 23 \\
CHStone    & N/A           & N/A                 & 12 \\
Linux      & 13,794        & 50                  & 50        \\
MiBench    & N/A           & N/A                 & 40 \\
NPB        & 22            & 50                  & 50               \\
OpenCV     & 342           & 50                  & 50               \\
TensorFlow & 1,885         & 50                  & 50           \\
\bottomrule
\end{tabular}

\end{table}

Additionally, we evaluate CompilerDream on a large-scale dataset named FormAI \cite{tihanyi2023formai}. FormAI comprises a vast collection of AI-generated C programs with diverse functionalities and coding styles. We filtered out codes that failed to compile into CompilerGym benchmarks, resulting in a test set of 109,016 programs.
 
Inspired by FormAI, we also constructed a dataset of 50 Objective-C programs generated by a Large Language Model to further evaluate CompilerDream's ability to generalize to different programming languages. 
Details can be found in Appendix \ref{app:compilergym_benchmarks}.

\subsection{Autotuning: CompilerGym Leaderboard}
\label{sec:search_exp}

\begin{table}[t]
\centering
\caption{
Autotuning results on cBench, i.e., the CompilerGym leaderboard \cite{compilergymleaderboard}, where learning-based methods report only inference time following leaderboard convention.}
\label{tab:leaderboard}
\begin{tabular}{lrc}
\toprule
\textbf{Method}                  & \textbf{Walltime}    & \begin{tabular}[c]{@{}l@{}}\textbf{Code Size}\\ \textbf{Reduction}\end{tabular} \\
\midrule
\textbf{\begin{tabular}[c]{@{}l@{}}CompilerDream\\ \quad+ Guided Search\end{tabular}}   & 60.8s     & \textbf{1.073$\times$}              \\
PPO + Guided Search     & 69.8s     & 1.070$\times$              \\
\textbf{CompilerDream}                & 2.9s     & 1.068$\times$              \\
Random Search ($t$=10800)& 10,512.4s & 1.062$\times$              \\
Random Search ($t$=3600)& 3,630.8s  & 1.061$\times$              \\
Greedy Search           & 169.2s    & 1.055$\times$              \\
GATv2 + DD-PPO          & 258.1s    & 1.047$\times$              \\
\bottomrule
\end{tabular}

\end{table}

We first validate CompilerDream as an autotuning method to demonstrate its ability to discover high-quality optimization sequences, supporting our goal of achieving superior code optimization.

\paragraph{\textbf{Implementation}} We target the CompilerGym leaderboard task \cite{compilergymleaderboard}, optimizing pass sequences for 23 cBench programs. Following the common setup adopted by other methods on the leaderboard, we train CompilerDream using our RL agent on all cBench programs except \textit{ghostscript}, which is excluded due to its large size that would significantly slow down training. The action space is set to the full action space of all 124 actions in LLVM. CompilerDream is trained for 25 hours, averaging about one hour per program—comparable to the 3600-second wall time of the random search algorithm. Beyond evaluating single-trial optimization performance, we test \textit{CompilerDream+Guided Search}, inspired by the leaderboard's leading approach. This method leverages the RL agent's policy $\pi_{\psi}(a_t|s_t)$ to guide random search, limiting search time to 1 minute per program (details in Appendix~\ref{app:search}).

\paragraph{\textbf{Results}}
As shown in Table \ref{tab:leaderboard}, our RL agent trained on world model simulations achieves an average $1.068\times$ code size reduction on cBench in a single trial, surpassing random search methods despite their longer execution times, even when accounting for CompilerDream's training time. This demonstrates the world model's ability to accurately simulate optimization processes and support effective agent learning. Furthermore, when combined with guided search, CompilerDream outperforms the top leaderboard method while using less wall time.

\subsection{General Value Prediction}
\label{sec:coreset_exp}

In this section, we demonstrate that CompilerDream's large-scale trained world model accurately simulates the optimization process of unseen programs, enabling the construction of an excellent value prediction agent that surpasses state-of-the-art methods.

\paragraph{\textbf{Implementation}}
To ensure a fair comparison of the value prediction capabilities, we adopt the setting of the state-of-the-art Coreset-NVP method \cite{liang2023rlcompopt}, which evaluates pass sequences from a fixed set of 50 distinct sequences (the \textit{core set}). We train our world model to build a value prediction agent (detailed in Section~\ref{sec:applications}) that predicts the value of applying sequences from the same core set. We follow the same search strategy as Coreset-NVP, enumerating the prefixes of roughly 3 or 4 sequences with the highest predicted value and selecting the best by compiler validation. Both methods are evaluated on four datasets distinct from their training sets, with an identical number of compiler validation calls to ensure a fair comparison. Additional details are provided in Appendix \ref{app:coreset}.

\begin{table}[t]
\centering
\caption{General value prediction results: The geometric mean of code size reduction achieved by the best sequences selected by different methods across 4 datasets. 
}
\label{tab:coreset}
\setlength{\tabcolsep}{1mm}
\begin{tabular}{lcccc}
\toprule
\multicolumn{1}{l}{\multirow{2}{*}{\textbf{Method}}} & \multicolumn{4}{c}{\textbf{Dataset}}        \\
\multicolumn{1}{c}{}                        & cBench & CHStone & MiBench & NPB   \\
\midrule

Coreset-NVP                                 & 1.028  & \textbf{1.101}  & 1.003   & 1.085 \\
\begin{tabular}[c]{@{}l@{}}\textbf{Coreset}\\ \quad \textbf{-CompilerDream}\end{tabular}                          & \textbf{1.038}          & \textbf{1.101} & \textbf{1.017} & \textbf{1.140} \\
\midrule
Coreset-Oracle                              & 1.041  & 1.106   & 1.020   & 1.159 \\
\bottomrule
\end{tabular}
\end{table}

\paragraph{\textbf{Results}}

Table \ref{tab:coreset} presents the geometric mean reduction achieved by different methods. Despite being tested on datasets distinct from the training set, CompilerDream surpasses Coreset-NVP on three datasets, demonstrating superior compiler simulation and generalization capabilities. On MiBench, CompilerDream performs nearly optimally compared to the Oracle baseline, which uses a brute-force search and serves as the problem's upper bound. Both our method and Coreset-NVP achieve an average reduction of $1.101$ on CHStone, close to the upper bound of $1.06$.

\subsection{General Code Optimization with RL}
\label{sec:main_exp}
We finally demonstrate the ability of CompilerDream's reinforcement learning agent, which can generate optimization sequences end-to-end in a single trial for a wide variety of programs unseen during training. This scenario mirrors real-world use cases where the optimization algorithm has limited time to compute the best sequence for arbitrary programs. Therefore, all methods in this section generate only one optimization sequence per program in the test datasets unless otherwise specified.

\begin{figure*}[ht]
    \centering
    \includegraphics[width=1.0\linewidth]{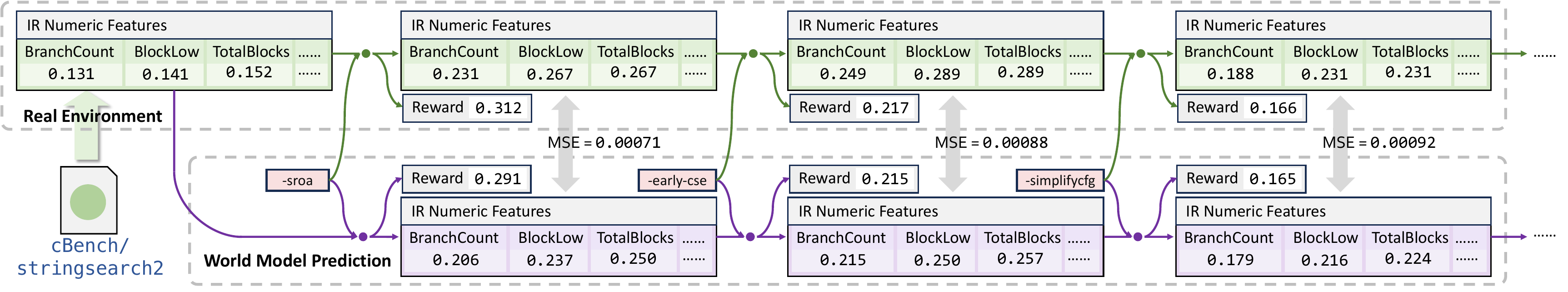}
    \caption{A comparison between a ground-truth code optimization trajectory and an imagined trajectory by a learned compiler world model. The learned world model accurately captures the variations of program features and optimization metrics.}
    \Description{This figure is the comparison between a ground-truth code optimization trajectory and an imagined trajectory by a learned compiler
world model. Both trajectories start from the ``cBench/stringsearch2" program and follow the same sequence of optimization passes (-sora, -early-cse, -simplifycfg, $\cdots$). Evaluating key metrics such as reward and intermediate representation (IR) features (BranchCount, BlockLow, TotalBlocks, $\cdots$), the world model's predictions closely align with the real environment. After the first three passes, the mean squared error (MSE) remains low at 0.00071, 0.00088, and 0.00092, demonstrating the model's accuracy in capturing optimization dynamics.}
    \label{fig:prediction}
\end{figure*}

\begin{figure*}[ht]
    \centering
    \includegraphics[width=1.0\linewidth]{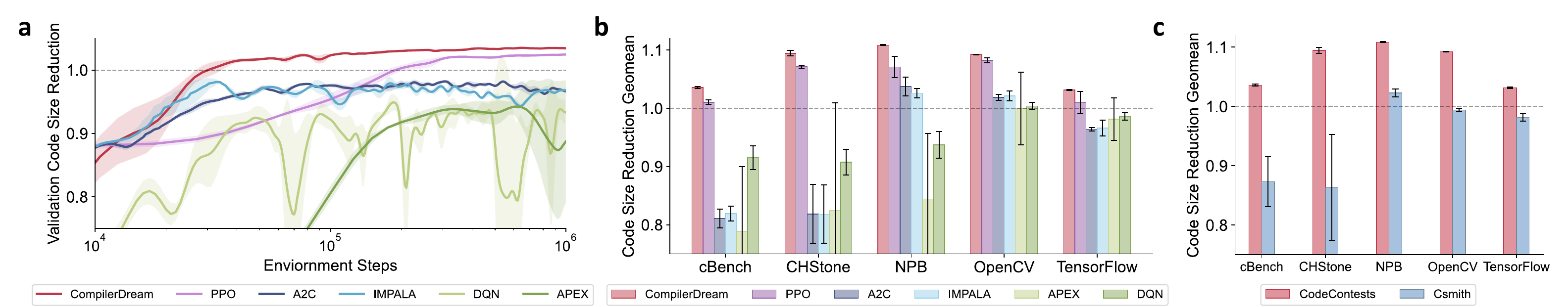}
    \caption{Analysis. Evaluations of different RL algorithms: (a) Learning curves of various RL algorithms, measured by the geometric mean of code size reduction on the CodeContests validation set. A Gaussian filter ($\sigma=2.0$) is applied to enhance the visualization of trends. (b) Generalization capabilities of different RL algorithms on various test datasets. Effect of training dataset: (c) Test performance of CompilerDream trained on CodeContests and Csmith. Bars indicate the standard deviation.}
    \Description{(a)The performance of CompilerDream steadily and consistently improves throughout training, and achieves a validation code size reduction of 1.02 after about $4 \times 10^4$ environment steps. In contrast, other methods (PPO, A2C, IMPALA, DQN, and APEX) all achieve reductions of less than 0.98 at the same stage. To reach a reduction of 1, PPO needs nearly $2 \times 10^5$ steps--roughly six times the steps required by CompilerDream. The remaining four methods exhibit unstable performance and fail to exceed a reduction of 1 even after  $10^6$ steps. (b) The generalization capabilities of different RL algorithms (CompilerDream, PPO, A2C, IMPALA, DQN, and APEX) on five test datasets unseen during training (cBench, CHStone, NPB, OpenCV, and TensorFlow). CompilerDream outperforms all other methods across all five datasets, demonstrating a clear advantage. PPO consistently ranks second, though with a noticeable performance gap compared to CompilerDream. The remaining four methods perform poorly, particularly on the cBench and CHStone datasets. (c) The impact of the training dataset is demonstrated by comparing CompilerDream trained on two datasets: CodeContests and Csmith. Across all five benchmarks (cBench, CHStone, NPB, OpenCV, and TensorFlow), CompilerDream trained on CodeContests consistently outperforms its counterpart trained on Csmith. Notably, the performance gap even reaches a geometric mean code size reduction difference of 0.2 on the CHStone benchmark.}
    \label{fig:ablation}
    \label{fig:dataset}
\end{figure*} 

\begin{figure}[hb]
    \centering
    \includegraphics[width=0.9\linewidth]{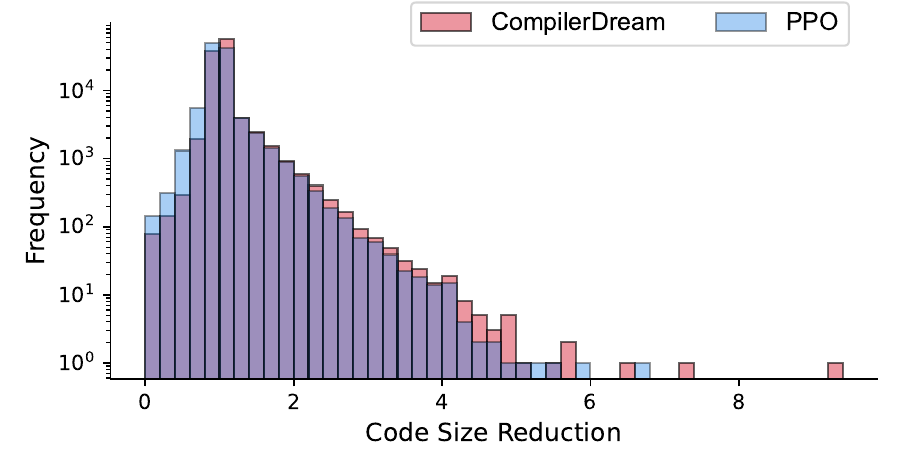}
    \caption{Histograms of large-scale evaluations comparing CompilerDream and PPO on the FormAI dataset. 
    }
    \Description{The performance of CompilerDream and PPO was evaluated on the large-scale FormAI test. For both methods, the code size reduction for most test cases (approximately $5 \times 10^4$) falls within the range of $0.8 \textasciitilde{} 1.2$. As the reduction magnitude increases or decreases beyond this range, the number of corresponding cases gradually decreases. Comparing CompilerDream with PPO, we observe that CompilerDream optimizes more cases with higher code size reduction while producing fewer cases with worsened results. Notably, CompilerDream achieves a reduction of more than nine times for some cases, surpassing all previous studies.}
    \label{fig:large_test}
\end{figure}

\paragraph{\textbf{Implementation}}

We train CompilerDream with its RL agent, and compare it with the following baselines: a random pass sequence, an autotuning approach using random search, LLVM's \texttt{-O0} and \texttt{-Oz} flag, and a state-of-the-art learning-based method \cite{huang2019autophase} using Proximal Policy Optimization (PPO) \cite{schulman2017proximal}. The random search conducts hundreds of trials within a time budget similar to our RL agent. The \texttt{-Oz} flag represents LLVM's highest level of code size optimization while the \texttt{-O0} flag represents no optimization. 
The learning-based method using PPO and CompilerDream are trained on the large-scale CodeContests dataset and zero-shot generalized to unseen test programs. 
Following the state-of-the-art approach, all methods use the reduced action space (Table~\ref{tab:obs_action}).

\paragraph{\textbf{Results}}
Figure \ref{fig:results} presents the code size reduction achieved by CompilerDream's RL agent, measured by the geometric mean of IR instruction count reduction. Without in-domain training, CompilerDream surpasses \texttt{-Oz} in all but two benchmarks in a single trial and consistently outperforms PPO, except on the BLAS datasets. It also outperforms random search on most datasets within a comparable time budget, except for Linux. The minimal performance differences across various methods on BLAS and Linux suggest these datasets are already highly optimized.

Moreover, CompilerDream's zero-shot generalization matches or surpasses in-domain training. This advantage is particularly evident in the NPB dataset, where data sparsity limits in-domain agents, yet CompilerDream achieves an additional 3\% code size reduction. Its robust generalization extends to new languages, as shown by results on the Fortran-based BLAS and NPB datasets. On the AI-generated Objective-C dataset, CompilerDream achieves an average code size reduction of 1.027$\times$, reaching up to 2.87$\times$ in certain test cases.

On the large-scale FormAI test set (Figure \ref{fig:large_test}), CompilerDream outperforms PPO, matching or outperforming \texttt{-Oz} on more programs and achieving higher optimization levels, demonstrating its superior and consistent performance.

\subsection{Analysis}
\label{sec:analysis}

\paragraph{\textbf{Comparison with model-free methods}}

We further evaluate the sample efficiency and zero-shot generalization abilities of our model-based CompilerDream (with RL agent) against various model-free counterparts, including PPO \cite{schulman2017proximal}, DQN \cite{mnih2015human}, A2C \cite{mnih2016asynchronous}, APEX \cite{horgan2018distributed}, and IMPALA \cite{espeholt2018impala, schulman2017proximal}. Figure~\ref{fig:ablation}a illustrates that while PPO is the strongest model-free baseline, our RL agent trained on world model simulation learns an order of magnitude faster with fewer compiler interactions. Figure~\ref{fig:ablation}b further highlights its superior generalization to unseen benchmarks, supporting our claim that world model-based agents better capture the dynamics of compiler optimization and enhance generalization to unseen datasets.

\paragraph{\textbf{Comparison with model-based methods}}
To demonstrate the necessity and effectiveness of learning a compiler world model with deep representations of the compiler. We compare our approach with a classical model-based RL method, MBPO \cite{janner2019trust}, under the RL agent setting described in Section~\ref{sec:main_exp}. MBPO adopts standard MLPs for its dynamics model and does not share a latent representation space with the actor or critic. For a fair comparison, we apply reward smoothing to MBPO using the same smoothing factor as in our method. The results are shown in Table~\ref{tab:compare_mbpo}. While MBPO outperforms the best model-free baseline, PPO, on some tasks, it consistently underperforms CompilerDream, except on MiBench, where the performance gap across all methods remains marginal. These results suggest that model-based RL alone is insufficient, and that both a more expressive world model and a shared latent space are crucial for better optimization performance and generalization to unseen programs.

\begin{table}[htb]
\centering  
\caption{Comparison between CompilerDream with MBPO and PPO for general code optimization.}
\label{tab:compare_mbpo}
\begin{tabular}{lccc}
\toprule
\textbf{Dataset}    & \textbf{PPO}   & \textbf{MBPO} & \textbf{CompilerDream} \\ \midrule
BLAS       & \textbf{0.993} & 0.988 	& 0.991 \\
cBench     & 1.010 & 1.020 	& \textbf{1.036} \\
CHStone    & 1.071 & 1.066 	& \textbf{1.094} \\
Linux      & 0.985 & \textbf{0.988}  & 0.986 \\
MiBench    & 1.000 & 0.996 	& \textbf{1.006} \\
NPB        & 1.071 & 1.069 	& \textbf{1.108} \\
OpenCV     & 1.082 & 0.998 	& \textbf{1.092} \\
TensorFlow & 1.010 & 0.996 	& \textbf{1.032} \\ \bottomrule
\end{tabular}
\end{table}

\paragraph{\textbf{Effect of reward smoothing}}
To evaluate the effectiveness of the reward smoothing technique described in Section~\ref{sec:model-based_method}, we compare the performance of the RL agent trained with the complete CompilerDream method against a variant without reward smoothing. Table~\ref{tab:reward_smoothing} presents these results alongside PPO for reference. The results demonstrate that reward smoothing consistently enhances optimization performance across all datasets and enables CompilerDream to outperform PPO on MiBench, Linux, and OpenCV.

\begin{table}[hbt]
\centering  
\caption{Comparison between CompilerDream with and without reward smoothing for general code optimization.}
\label{tab:reward_smoothing}
\setlength\tabcolsep{3pt}
\begin{tabular}{lccc}
\toprule
\textbf{Dataset}    & \textbf{PPO}   & \textbf{\begin{tabular}[c]{@{}c@{}}Ours w/o \\ Reward Smoothing\end{tabular}} & \textbf{\begin{tabular}[c]{@{}c@{}}Ours w/ \\ Reward Smoothing\end{tabular}} \\ \midrule
BLAS       & \textbf{0.993} & 0.987                                                          & 0.991                                                           \\
cBench     & 1.010          & 1.021                                                          & \textbf{1.036}                                                  \\
CHStone    & 1.071          & 1.076                                                          & \textbf{1.094}                                                  \\
MiBench    & 1.000          & 0.995                                                          & \textbf{1.006}                                                  \\
NPB        & 1.071          & 1.093                                                          & \textbf{1.108}                                                  \\
Linux      & 0.985          & 0.980                                                          & \textbf{0.986}                                                  \\
OpenCV     & 1.082          & 1.080                                                          & \textbf{1.092}                                                  \\
TensorFlow & 1.010          & 1.022                                                          & \textbf{1.032}                                                  \\ \bottomrule
\end{tabular}
\end{table}

\paragraph{\textbf{Effect of training dataset}}

To assess the impact of the CodeContests dataset on the generalization ability, we compared it with the commonly used Csmith \cite{yang2011finding} dataset, a large LLVM IR dataset generated by rules. We train CompilerDream with RL agent on both datasets and the results shown in Figure \ref{fig:dataset} indicate that CompilerDream trained on CodeContests significantly outperforms the Csmith-trained version across all five test datasets. This advantage is particularly evident in the manually curated cBench \cite{fursin2007midatasets} and CHStone \cite{hara2008chstone} datasets, demonstrating that the CodeContests-trained model can generalize more effectively to human-written programs.

\paragraph{\textbf{Program showcase}}

In Figure~\ref{fig:prediction}, we display a predicted optimization trajectory for an unseen program from cBench, as forecasted by our learned compiler world model. The model successfully forecasts numeric features of future IR, including the counts of branches and blocks, alongside future rewards that signify optimization outcomes. This instance exemplifies the capability of our learned compiler world model to serve as a viable alternative for a real compiler environment in training code optimization agents.

\section{Discussion}

We aim to address the major challenge of generalization in learning-based code optimization by introducing the CompilerDream approach, which leverages the simulation and generalization capabilities of a world model. By incorporating a reward smoothing technique and a large-scale training program dataset, we enable effective world model training for compiler optimization tasks. Our method supports two types of optimization agents: a value prediction agent and an RL agent. Experimental results demonstrate that CompilerDream achieves superior optimization performance across diverse problem scenarios and program datasets, surpassing built-in compiler optimization flags and state-of-the-art methods.

\paragraph{\textbf{Limitation}}
Although our method can naturally extend to optimization objectives such as execution time and object file size reduction, we focus solely on code size optimization for scalability and stability, as discussed in Section~\ref{sec:eval}. Execution time measurements often exhibit high variance and biases due to hardware and operating system states, introducing significant noise that additional sampling cannot mitigate. Furthermore, existing open-source frameworks, such as CompilerGym, lack robust support for accurate runtime measurement, making it challenging to obtain reliable signals for effective learning. As a result, we only focus on code size reduction for this paper. 

\paragraph{\textbf{Future Works}}
There is substantial scope for further exploration, including expansion of the training dataset, scaling up the compiler world model, optimizing multiple objectives like execution time, and enriching feature and action spaces with deeper expert knowledge or large language models.
Additionally, our approach could support more types of optimization agents, such as search agents leveraging advanced tree search algorithms \cite{schrittwieser2020mastering} based on world model simulations.

\section{Acknowledgments}
This work was supported by the National Natural Science Foundation of China (62021002).
\bibliographystyle{ACM-Reference-Format}
\bibliography{compilerdream}

@inproceedings{brauckmann2020compiler,
  title={Compiler-based graph representations for deep learning models of code},
  author={Brauckmann, Alexander and Goens, Andr{\'e}s and Ertel, Sebastian and Castrillon, Jeronimo},
  booktitle={Proceedings of the 29th International Conference on Compiler Construction},
  pages={201--211},
  year={2020}
}

@article{van2008visualizing,
  title={Visualizing data using t-SNE.},
  author={Van der Maaten, Laurens and Hinton, Geoffrey},
  journal={Journal of machine learning research},
  volume={9},
  number={11},
  year={2008}
}

@article{filho2018yet,
  title={Yet another intelligent code-generating system: A flexible and low-cost solution},
  author={Filho, Jo{\~a}o Fabr{\'\i}cio and Rodriguez, Luis Gustavo Araujo and da Silva, Anderson Faustino},
  journal={Journal of Computer Science and Technology},
  volume={33},
  pages={940--965},
  year={2018},
  publisher={Springer}
}

@article{ansel2014opentuner,
  title={Opentuner: An extensible framework for program autotuning},
  author={Ansel, Jason and Kamil, Shoaib and Veeramachaneni, Kalyan and Ragan-Kelley, Jonathan and Bosboom, Jeffrey and O'Reilly, Una-May and Amarasinghe, Saman},
  journal={Proceedings of the 23rd international conference on Parallel architectures and compilation (MEMSYS '24)},
  pages={303--316},
  year={2014}
}

@article{alon2019code2vec,
  title={code2vec: Learning distributed representations of code},
  author={Alon, Uri and Zilberstein, Meital and Levy, Omer and Yahav, Eran},
  journal={Proceedings of the ACM on Programming Languages},
  volume={3},
  number={POPL},
  pages={1--29},
  year={2019},
  publisher={ACM New York, NY, USA}
}

@article{venkatakeerthy2020ir2vec,
  title={Ir2vec: Llvm ir based scalable program embeddings},
  author={VenkataKeerthy, S and Aggarwal, Rohit and Jain, Shalini and Desarkar, Maunendra Sankar and Upadrasta, Ramakrishna and Srikant, YN},
  journal={ACM Transactions on Architecture and Code Optimization (TACO '20)},
  volume={17},
  number={4},
  pages={1--27},
  year={2020},
  publisher={ACM New York, NY, USA}
}

@inproceedings{ben2018neural,
    author = {Ben-Nun, Tal and Jakobovits, Alice Shoshana and Hoefler, Torsten},
    title = {Neural code comprehension: a learnable representation of code semantics},
    year = {2018},
    booktitle = {Proceedings of the 32nd International Conference on Neural Information Processing Systems},
    pages = {3589–3601},
    series = {NeurIPS '18}
}

@article{mikolov2013distributed,
  title={Distributed representations of words and phrases and their compositionality},
  author={Mikolov, Tomas and Sutskever, Ilya and Chen, Kai and Corrado, Greg S and Dean, Jeff},
  journal={Advances in Neural Information Processing Systems (NeurIPS '13)},
  year={2013}
}

@article{kulkarni2012mitigating,
  title={Mitigating the compiler optimization phase-ordering problem using machine learning},
  author={Kulkarni, Sameer and Cavazos, John},
  journal={Proceedings of the ACM international conference on Object oriented programming systems languages and applications},
  pages={147--162},
  year={2012}
}

@article{magni2014automatic,
  title={Automatic optimization of thread-coarsening for graphics processors},
  author={Magni, Alberto and Dubach, Christophe and O'Boyle, Michael},
  journal={Proceedings of the 23rd international conference on Parallel architectures and compilation},
  pages={455--466},
  year={2014}
}

@inproceedings{brauckmann2021reinforcement,
author = {Brauckmann, Alexander and Goens, Andr\'{e}s and Castrillon, Jeronimo},
title = {PolyGym: Polyhedral Optimizations as an Environment for Reinforcement Learning},
year = {2024},
isbn = {9781665442787},
publisher = {IEEE Press},
booktitle = {Proceedings of the 30th International Conference on Parallel Architectures and Compilation Techniques},
pages = {17–29},
numpages = {13},
location = {Atlanta, GA, USA},
series = {PACT '21}
}

@inproceedings{mirhoseini2017device,
  title={Device placement optimization with reinforcement learning},
  author={Mirhoseini, Azalia and Pham, Hieu and Le, Quoc V and Steiner, Benoit and Larsen, Rasmus and Zhou, Yuefeng and Kumar, Naveen and Norouzi, Mohammad and Bengio, Samy and Dean, Jeff},
  booktitle={International Conference on Machine Learning (ICML '17)},
  pages={2430--2439},
  year={2017},
}

@inproceedings{haj2020neurovectorizer,
  title={Neurovectorizer: End-to-end vectorization with deep reinforcement learning},
  author={Haj-Ali, Ameer and Ahmed, Nesreen K and Willke, Ted and Shao, Yakun Sophia and Asanovic, Krste and Stoica, Ion},
  booktitle={International Symposium on Code Generation and Optimization},
  pages={242--255},
  year={2020},
  publisher = {IEEE Press},
  series = {CGO '20}
}

@inproceedings{luk2009qilin,
  author={Luk, Chi-Keung and Hong, Sunpyo and Kim, Hyesoon},
  booktitle={2009 42nd Annual IEEE/ACM International Symposium on Microarchitecture (MICRO '09)}, 
  title={Qilin: Exploiting parallelism on heterogeneous multiprocessors with adaptive mapping}, 
  year={2009},
  volume={},
  number={},
  pages={45-55},
  doi={}}

@article{stephenson2003meta,
  title={Meta optimization: Improving compiler heuristics with machine learning},
  author={Stephenson, Mark and Amarasinghe, Saman and Martin, Martin and O'Reilly, Una-May},
  journal={ACM sigplan notices},
  volume={38},
  number={5},
  pages={77--90},
  year={2003},
  publisher={ACM New York, NY, USA}
}

@inproceedings{datta2008stencil,
  author = {Datta, Kaushik and Murphy, Mark and Volkov, Vasily and Williams, Samuel and Carter, Jonathan and Oliker, Leonid and Patterson, David and Shalf, John and Yelick, Katherine},
title = {Stencil computation optimization and auto-tuning on state-of-the-art multicore architectures},
year = {2008},
isbn = {9781424428359},
publisher = {IEEE Press},
booktitle = {Proceedings of the 2008 ACM/IEEE Conference on Supercomputing},
articleno = {4},
numpages = {12},
location = {Austin, Texas},
series = {SC '08}
}

@article{kirillov2023segment,
    author    = {Kirillov, Alexander and Mintun, Eric and Ravi, Nikhila and Mao, Hanzi and Rolland, Chloe and Gustafson, Laura and Xiao, Tete and Whitehead, Spencer and Berg, Alexander C. and Lo, Wan-Yen and Dollar, Piotr and Girshick, Ross},
    title     = {Segment Anything},
    journal = {Proceedings of the IEEE/CVF International Conference on Computer Vision (ICCV)},
    month     = {October},
    year      = {2023},
    pages     = {4015-4026}
}

@article{mnih2015human,
  title={Human-level control through deep reinforcement learning},
  author={Mnih, Volodymyr and Kavukcuoglu, Koray and Silver, David and Rusu, Andrei A and Veness, Joel and Bellemare, Marc G and Graves, Alex and Riedmiller, Martin and Fidjeland, Andreas K and Ostrovski, Georg and others},
  journal={Nature},
  volume={518},
  number={7540},
  pages={529--533},
  year={2015},
  publisher={Nature Publishing Group}
}

@Eprint{schulman2017proximal,
      title={Proximal Policy Optimization Algorithms}, 
      author={John Schulman and Filip Wolski and Prafulla Dhariwal and Alec Radford and Oleg Klimov},
      year={2017},
      eprint={1707.06347},
      archivePrefix={arXiv},
      primaryClass={cs.LG},
      url={https://arxiv.org/abs/1707.06347}, 
}

@inproceedings{espeholt2018impala,
  title={Impala: Scalable distributed deep-rl with importance weighted actor-learner architectures},
  author={Espeholt, Lasse and Soyer, Hubert and Munos, Remi and Simonyan, Karen and Mnih, Vlad and Ward, Tom and Doron, Yotam and Firoiu, Vlad and Harley, Tim and Dunning, Iain and others},
  booktitle={International Conference on Machine Learning (ICML '18)},
  pages={1407--1416},
  year={2018},
}

@inproceedings{mnih2016asynchronous,
  title={Asynchronous methods for deep reinforcement learning},
  author={Mnih, Volodymyr and Badia, Adria Puigdomenech and Mirza, Mehdi and Graves, Alex and Lillicrap, Timothy and Harley, Tim and Silver, David and Kavukcuoglu, Koray},
  booktitle={International Conference on Machine Learning (ICML '16)},
  pages={1928--1937},
  year={2016},
}

@Eprint{trofin2021mlgo,
      title={MLGO: a Machine Learning Guided Compiler Optimizations Framework}, 
      author={Mircea Trofin and Yundi Qian and Eugene Brevdo and Zinan Lin and Krzysztof Choromanski and David Li},
      year={2021},
      eprint={2101.04808},
      archivePrefix={arXiv},
      primaryClass={cs.PL},
      url={https://arxiv.org/abs/2101.04808}, 
}

@book{sutton2018reinforcement,
  title={Reinforcement learning: An introduction},
  author={Sutton, Richard S and Barto, Andrew G},
  year={2018},
  publisher={MIT press}
}

@inproceedings{lee2023dreamsmooth,
    title={DreamSmooth: Improving Model-based Reinforcement Learning via Reward Smoothing},
    author={Vint Lee and Pieter Abbeel and Youngwoon Lee},
    booktitle={International Conference on Learning Representations (ICLR '24)},
    year={2024},
}

@inproceedings{mazoure2021cross,
title={Cross-Trajectory Representation Learning for Zero-Shot Generalization in {RL}},
author={Bogdan Mazoure and Ahmed M Ahmed and R Devon Hjelm and Andrey Kolobov and Patrick MacAlpine},
booktitle={International Conference on Learning Representations (ICLR '22)},
year={2022},
}

@article{lawson1979basic,
  title={Basic linear algebra subprograms for Fortran usage},
  author={Lawson, Chuck L and Hanson, Richard J. and Kincaid, David R and Krogh, Fred T.},
  journal={ACM Transactions on Mathematical Software (TOMS '79)},
  volume={5},
  number={3},
  pages={308--323},
  year={1979},
  publisher={ACM New York, NY, USA}
}

@inproceedings{fursin2007midatasets,
  title={Midatasets: Creating the conditions for a more realistic evaluation of iterative optimization},
  author={Fursin, Grigori and Cavazos, John and O'Boyle, Michael and Temam, Olivier},
  booktitle={International conference on high-performance embedded architectures and compilers},
  pages={245--260},
  year={2007},
  series = {HiPEAC '07},
  publisher = {Springer-Verlag},
  address = {Berlin, Heidelberg},
}

@article{hara2008chstone,
   author={Yuko Hara and Hiroyuki Tomiyama and Shinya Honda and Hiroaki Takada and Katsuya Ishii},
  booktitle={2008 IEEE International Symposium on Circuits and Systems (ISCAS '08)}, 
  title={CHStone: A benchmark program suite for practical C-based high-level synthesis}, 
  year={2008},
  volume={},
  number={},
  pages={1192-1195},

}

@inproceedings{guthaus2001mibench,
  title={MiBench: A free, commercially representative embedded benchmark suite},
  author={Guthaus, Matthew R and Ringenberg, Jeffrey S and Ernst, Dan and Austin, Todd M and Mudge, Trevor and Brown, Richard B},
  booktitle={Proceedings of the fourth annual IEEE international workshop on workload characterization},
  pages={3--14},
  year={2001},
  series = {WWC '01}
}

@techreport{bailey1995parallel,
  title={The {NAS} parallel benchmarks 2.0},
  author={Bailey, David and Harris, Tim and Saphir, William and Van Der Wijngaart, Rob and Woo, Alex and Yarrow, Maurice},
  year={1995},
  institution={Technical Report NAS-95-020, NASA Ames Research Center}
}

@misc{compilergymleaderboard,
  title        = "CompilerGym Leaderboard",
  author       = "{Facebook}",
  howpublished = "\url{https://github.com/facebookresearch/CompilerGym?tab=readme-ov-file\#leaderboards}",
  year         = 2022,
  note         = "Accessed: 2024-8-09"
}

@article{williams1992simple,
  title={Simple statistical gradient-following algorithms for connectionist reinforcement learning},
  author={Williams, Ronald J},
  journal={Machine learning},
  volume={8},
  pages={229--256},
  year={1992},
  publisher={Springer}
}

@inproceedings{yang2011finding,
  title={Finding and understanding bugs in C compilers},
  author={Yang, Xuejun and Chen, Yang and Eide, Eric and Regehr, John},
  booktitle={Proceedings of the 32nd ACM SIGPLAN conference on Programming language design and implementation},
  pages={283--294},
  year={2011},
  series = {PLDI '18}
}

@inproceedings{da2021anghabench,
  author = {da Silva, Anderson Faustino and Kind, Bruno Conde and de Souza Magalh\~{a}es, Jos\'{e} Wesley and Rocha, Jer\^{o}nimo Nunes and Guimar\~{a}es, Breno Campos Ferreira and Pereira, Fernando Magno Quint\~{a}o},
title = {AnghaBench: a suite with one million compilable C benchmarks for code-size reduction},
year = {2021},
isbn = {9781728186139},
publisher = {IEEE Press},
booktitle = {Proceedings of the 2021 IEEE/ACM International Symposium on Code Generation and Optimization},
pages = {378–390},
numpages = {13},
series = {CGO '21}
}

@inproceedings{abadi2016tensorflow,
  author={Abadi, Mart{\'\i}n and Barham, Paul and Chen, Jianmin and Chen, Zhifeng and Davis, Andy and Dean, Jeffrey and Devin, Matthieu and Ghemawat, Sanjay and Irving, Geoffrey and Isard, Michael and others},
    title = {{TensorFlow}: A System for {Large-Scale} Machine Learning},
    booktitle = {12th USENIX Symposium on Operating Systems Design and Implementation (OSDI 16)},
    year = {2016},
    isbn = {978-1-931971-33-1},
    address = {Savannah, GA},
    pages = {265--283},
    publisher = {USENIX Association},
    month = nov
}

@INPROCEEDINGS{culjak2012brief,
  author={Culjak, Ivan and Abram, David and Pribanic, Tomislav and Dzapo, Hrvoje and Cifrek, Mario},
  booktitle={2012 Proceedings of the 35th International Convention MIPRO}, 
  title={A brief introduction to OpenCV}, 
  year={2012},
  volume={},
  number={},
  pages={1725--1730},
  doi={},
  series={MIPRO '12}
}

@inproceedings{cummins2022compilergym,
author = {Cummins, Chris and Wasti, Bram and Guo, Jiadong and Cui, Brandon and Ansel, Jason and Gomez, Sahir and Jain, Somya and Liu, Jia and Teytaud, Olivier and Steiner, Benoit and Tian, Yuandong and Leather, Hugh},
title = {CompilerGym: robust, performant compiler optimization environments for AI research},
year = {2022},
publisher = {IEEE Press},
booktitle = {Proceedings of the 20th IEEE/ACM International Symposium on Code Generation and Optimization},
pages = {92–105},
numpages = {14},
series = {CGO '22}
}

@inproceedings{tihanyi2023formai,
  title={The formai dataset: Generative ai in software security through the lens of formal verification},
  author={Tihanyi, Norbert and Bisztray, Tamas and Jain, Ridhi and Ferrag, Mohamed Amine and Cordeiro, Lucas C and Mavroeidis, Vasileios},
  booktitle={Proceedings of the 19th International Conference on Predictive Models and Data Analytics in Software Engineering},
  pages={33--43},
  year={2023},
  series = {PROMISE '23}
}

@article{li2022competition,
  title={Competition-level code generation with alphacode},
  author={Li, Yujia and Choi, David and Chung, Junyoung and Kushman, Nate and Schrittwieser, Julian and Leblond, R{\'e}mi and Eccles, Tom and Keeling, James and Gimeno, Felix and Dal Lago, Agustin and others},
  journal={Science},
  volume={378},
  number={6624},
  pages={1092--1097},
  year={2022},
  publisher={American Association for the Advancement of Science}
}

@inproceedings{fursin2008milepost,
  title={MILEPOST GCC: machine learning based research compiler},
  author={Fursin, Grigori and Miranda, Cupertino and Temam, Olivier and Namolaru, Mircea and Zaks, Ayal and Mendelson, Bilha and Bonilla, Edwin and Thomson, John and Leather, Hugh and Williams, Chris and others},
  booktitle={GCC summit},
ADDRESS = {Ottawa, Canada},
MONTH = Jun,
  year={2008}
}

@article{calder1997evidence,
  title={Evidence-based static branch prediction using machine learning},
  author={Calder, Brad and Grunwald, Dirk and Jones, Michael and Lindsay, Donald and Martin, James and Mozer, Michael and Zorn, Benjamin},
  journal={ACM Transactions on Programming Languages and Systems (TOPLAS '97)},
  volume={19},
  number={1},
  pages={188--222},
  year={1997},
  publisher={ACM}
}

@inproceedings{radford2021learning,
  title={Learning transferable visual models from natural language supervision},
  author={Radford, Alec and Kim, Jong Wook and Hallacy, Chris and Ramesh, Aditya and Goh, Gabriel and Agarwal, Sandhini and Sastry, Girish and Askell, Amanda and Mishkin, Pamela and Clark, Jack and others},
  booktitle={International Conference on Machine Learning (ICML '21)},
  pages={8748--8763},
  year={2021},
}

@inproceedings{devlin2018bert,
  title={BERT: Pre-training of Deep Bidirectional Transformers for Language Understanding},
  author={Jacob Devlin and Ming-Wei Chang and Kenton Lee and Kristina Toutanova},
  booktitle={Proceedings of the 2019 Conference of the North American Chapter of
                  the Association for Computational Linguistics: Human Language Technologies,
                  (NAACL-HLT ’19)},
  year={2019},
 pages        = {4171--4186},
  publisher    = {Association for Computational Linguistics},
}

@inproceedings{brown2020language,
    title={Language models are few-shot learners},
    author={Brown, Tom and Mann, Benjamin and Ryder, Nick and Subbiah, Melanie and Kaplan, Jared D and Dhariwal, Prafulla and Neelakantan, Arvind and Shyam, Pranav and Sastry, Girish and Askell, Amanda and others},
    booktitle = {Proceedings of the 34th International Conference on Neural Information Processing Systems},
    articleno = {159},
    numpages = {25},
    series = {NeurIPS '20},
    year = {2020},
}

@inproceedings{triantafyllis2003compiler,
  title={Compiler optimization-space exploration},
  author={Triantafyllis, Spyridon and Vachharajani, Manish and Vachharajani, Neil and August, David I},
  booktitle={International Symposium on Code Generation and Optimization},
  pages={204--215},
  year={2003},
  publisher = {IEEE Press},
  series = {CGO '03}
}

@inproceedings{georgiou2018less,
  title={Less is more: Exploiting the standard compiler optimization levels for better performance and energy consumption},
  author={Georgiou, Kyriakos and Blackmore, Craig and Xavier-de-Souza, Samuel and Eder, Kerstin},
  booktitle={Proceedings of the 21st International Workshop on Software and Compilers for Embedded Systems},
  pages={35--42},
  year={2018},
  series = {SCOPES '18}
}

@inproceedings{bodin1998iterative,
  TITLE = {{Iterative compilation in a non-linear optimisation space}},
  AUTHOR = {Bodin, Fran{\c c}ois and Kisuki, Toru and Knijnenburg, Peter and O' Boyle, Mike and Rohou, Erven},
  BOOKTITLE = {{Workshop on Profile and Feedback-Directed Compilation}},
  ADDRESS = {Paris, France},
  YEAR = {1998},
  MONTH = Oct,
  HAL_ID = {inria-00475919},
  HAL_VERSION = {v1},
}

@inproceedings{anand2021procedural,
  title={Procedural generalization by planning with self-supervised world models},
  author={Anand, Ankesh and Walker, Jacob C and Li, Yazhe and V{\'e}rtes, Eszter and Schrittwieser, Julian and Ozair, Sherjil and Weber, Theophane and Hamrick, Jessica B},
  booktitle={Proceedings of the 10th International Conference on Learning Representations (ICLR '22)},
  year={2022}
}

@inproceedings{shahzad2022reinforcement,
  title={Reinforcement Learning Strategies for Compiler Optimization in High Level Synthesis},
  author={Shahzad, Hafsah and Sanaullah, Ahmed and Arora, Sanjay and Munafo, Robert and Yao, Xiteng and Drepper, Ulrich and Herbordt, Martin},
  booktitle={Proceedings of the 2022 IEEE/ACM 8th Workshop on the LLVM Compiler Infrastructure in HPC (LLVM-HPC ’22) and Workshop on Hierarchical Parallelism for Exascale Computing (HiPar ’22)},
  pages={13--22},
  year={2022},
}

@inproceedings{mammadli2020static,
  title={Static Neural Compiler Optimization via Deep Reinforcement Learning},
  author={Mammadli, Rahim and Jannesari, Ali and Wolf, Felix},
  booktitle={Proceedings of the 2020 IEEE/ACM 6th Workshop on the LLVM Compiler Infrastructure in HPC (LLVM-HPC ’20) and Workshop on Hierarchical Parallelism for Exascale Computing (HiPar ’20)},
  pages={1--11},
  year={2020},
}

@inproceedings{leather2020machine,
  title={Machine learning in compilers: Past, present and future},
  author={Leather, Hugh and Cummins, Chris},
  booktitle={2020 Forum for Specification and Design Languages (FDL)},
  pages={1--8},
  year={2020},
  organization={IEEE}
}

@article{cummins2023large,
  title={Large language models for compiler optimization},
  author={Cummins, Chris and Seeker, Volker and Grubisic, Dejan and Elhoushi, Mostafa and Liang, Youwei and Roziere, Baptiste and Gehring, Jonas and Gloeckle, Fabian and Hazelwood, Kim and Synnaeve, Gabriel and others},
  journal={arXiv preprint arXiv:2309.07062},
  year={2023}
}

@InProceedings{liang2023rlcompopt,
  title={Learning Compiler Pass Orders using Coreset and Normalized Value Prediction},
  author={Liang, Youwei and Stone, Kevin and Shameli, Ali and Cummins, Chris and Elhoushi, Mostafa and Guo, Jiadong and Steiner, Benoit and Yang, Xiaomeng and Xie, Pengtao and Leather, Hugh and Tian, Yuandong},
  year={2023},
  booktitle={Proceedings of the 40th International Conference on Machine Learning (ICML '23)}

}

@inproceedings{jain2022poset,
  title={{POSET-RL}: Phase ordering for optimizing size and execution time using reinforcement learning},
  author={Jain, Shalini and Andaluri, Yashas and VenkataKeerthy, S and Upadrasta, Ramakrishna},
  booktitle={2022 IEEE International Symposium on Performance Analysis of Systems and Software (ISPASS '22)},
  pages={121--131},
  year={2022},
  organization={IEEE}
}

@inproceedings{lattner2004llvm,
  title={LLVM: A compilation framework for lifelong program analysis \& transformation},
  author={Lattner, Chris and Adve, Vikram},
  booktitle={International Symposium on Code Generation and Optimization},
  pages={75--86},
  year={2004},
  publisher = {IEEE Press},
  series = {CGO '04}
}

@article{cummins2017end,
  title={End-to-end deep learning of optimization heuristics},
  author={Cummins, Chris and Petoumenos, Pavlos and Wang, Zheng and Leather, Hugh},
  journal={International Conference on Parallel Architectures and Compilation Techniques (PACT)},
  pages={219--232},
  year={2017},
  organization={IEEE}
}

@article{alon2018general,
  title={A general path-based representation for predicting program properties},
  author={Alon, Uri and Zilberstein, Meital and Levy, Omer and Yahav, Eran},
  journal={ACM SIGPLAN Notices},
  volume={53},
  number={4},
  pages={404--419},
  year={2018},
  publisher={ACM New York, NY, USA}
}

@inproceedings{huang2019autophase,
  author       = {Ameer Haj{-}Ali and
                  Qijing (Jenny) Huang and
                  William S. Moses and
                  John Xiang and
                  Krste Asanovic and
                  John Wawrzynek and
                  Ion Stoica},
  title        = {AutoPhase: Juggling {HLS} Phase Orderings in Random Forests with Deep
                  Reinforcement Learning},
  booktitle = {Proceedings of Machine Learning and Systems},
  year         = {2020},
   pages = {70--81},
  series = {MLSys '20}
}

@inproceedings{cummins2020programl,
  title={ProGraML: A Graph-based Program Representation for Data Flow Analysis and Compiler Optimizations},
  author={Cummins, Chris and Fisches, Zacharias V. and Ben-Nun, Tal and Hoefler, Torsten and O'Boyle, Michael F P and Leather, Hugh},
  booktitle={Proceedings of the 38th International Conference on Machine Learning (ICML '21)},
  pages={2244--2253},
  year={2021},
  month={18--24 Jul},
}

@article{hafner2023mastering,
  title={Mastering diverse domains through world models},
  author={Hafner, Danijar and Pasukonis, Jurgis and Ba, Jimmy and Lillicrap, Timothy},
  journal={arXiv preprint arXiv:2301.04104},
  year={2023}
}

@inproceedings{hafnermastering,
  title={Mastering Atari with Discrete World Models},
  author={Hafner, Danijar and Lillicrap, Timothy P and Norouzi, Mohammad and Ba, Jimmy},
  booktitle={International Conference on Learning Representations (ICLR '21)},
  year={2021},
  
}

@inproceedings{horgan2018distributed,
  title={Distributed Prioritized Experience Replay},
  author={Dan Horgan and John Quan and David Budden and Gabriel Barth-Maron and Matteo Hessel and Hado van Hasselt and David Silver},
  booktitle={International Conference on Learning Representations (ICLR '18)},
  year={2018},
}

@article{dqn,
  title={Human-level control through deep reinforcement learning},
  author={Mnih, Volodymyr and Kavukcuoglu, Koray and Silver, David and Rusu, Andrei A and Veness, Joel and Bellemare, Marc G and Graves, Alex and Riedmiller, Martin and Fidjeland, Andreas K and Ostrovski, Georg and others},
  journal={Nature},
  volume={518},
  number={7540},
  pages={529},
  year={2015},
  publisher={Nature Publishing Group}
}

@inproceedings{liang2018rllib,
    Author = {Eric Liang and
              Richard Liaw and
              Robert Nishihara and
              Philipp Moritz and
              Roy Fox and
              Ken Goldberg and
              Joseph E. Gonzalez and
              Michael I. Jordan and
              Ion Stoica},
    Title = {{RLlib}: Abstractions for Distributed Reinforcement Learning},
    booktitle = {International Conference on Machine Learning ({ICML '18})},
    Year = {2018},
    pages =  {3053--3062},  
    volume =  {80}
}

@inproceedings{ma2024harmonydream,
  title={HarmonyDream: Task Harmonization Inside World Models},
  author={Ma, Haoyu and Wu, Jialong and Feng, Ningya and Xiao, Chenjun and Li, Dong and Jianye, HAO and Wang, Jianmin and Long, Mingsheng},
  booktitle={International Conference on Machine Learning ({ICML '24})},
  year={2024}
}

@Eprint{haarnoja2018soft,
  title={Soft Actor-Critic Algorithms and Applications}, 
      author={Tuomas Haarnoja and Aurick Zhou and Kristian Hartikainen and George Tucker and Sehoon Ha and Jie Tan and Vikash Kumar and Henry Zhu and Abhishek Gupta and Pieter Abbeel and Sergey Levine},
      year={2019},
      eprint={1812.05905},
      archivePrefix={arXiv},
      primaryClass={cs.LG},
      url={https://arxiv.org/abs/1812.05905}, 
}

@inproceedings{ha2018recurrent,
  title={Recurrent world models facilitate policy evolution},
  author={Ha, David and Schmidhuber, J{\"u}rgen},
  booktitle={Proceedings of the 32nd International Conference on Neural Information Processing Systems},
  year={2018},
pages = {2455–2467},
   series = {NeurIPS '18},
}

@article{lecun2022path,
  title={A path towards autonomous machine intelligence version 0.9. 2, 2022-06-27},
  author={LeCun, Yann},
  journal={Open Review},
  volume={62},
  number={1},
  pages={1--62},
  year={2022},
  url={https://openreview.net/pdf?id=BZ5a1r-kVsf}
}

@inproceedings{janner2019trust,
  title={When to trust your model: Model-based policy optimization},
  author={Janner, Michael and Fu, Justin and Zhang, Marvin and Levine, Sergey},
  booktitle={Proceedings of the 33rd International Conference on Neural Information Processing Systems},
  articleno = {1122},
  series={NeurIPS ’19},
  year={2019}
}

@Eprint{hafner2019dream,
      title={Dream to Control: Learning Behaviors by Latent Imagination}, 
      author={Danijar Hafner and Timothy Lillicrap and Jimmy Ba and Mohammad Norouzi},
      year={2020},
      eprint={1912.01603},
      archivePrefix={arXiv},
      primaryClass={cs.LG},
      url={https://arxiv.org/abs/1912.01603}, 
}

@article{schrittwieser2020mastering,
  title={Mastering atari, go, chess and shogi by planning with a learned model},
  author={Schrittwieser, Julian and Antonoglou, Ioannis and Hubert, Thomas and Simonyan, Karen and Sifre, Laurent and Schmitt, Simon and Guez, Arthur and Lockhart, Edward and Hassabis, Demis and Graepel, Thore and others},
  journal={Nature},
  volume={588},
  number={7839},
  pages={604--609},
  year={2020},
  publisher={Nature Publishing Group UK London}
}

\appendix

\section{Extended Related Work}

Here we provide an overview of several topics in compiler optimization that our work did not focus on but are nonetheless important:

\paragraph{Feature Extraction} Machine learning techniques introduced in Section \ref{related_work} require crafting high-quality features that capture the important characteristics of programs, a process known as feature engineering. The most prevalent feature vectors are based on the frequencies of various types of instructions or patterns within the programs \cite{magni2014automatic, huang2019autophase}, designed by expert intuitions. Numerous studies have aimed to reduce the cost of feature design. Following the success of word2vec embeddings in natural language processing \cite{mikolov2013distributed}, methods like code2vec \cite{alon2019code2vec}, inst2vec \cite{ben2018neural}, and IR2vec \cite{venkatakeerthy2020ir2vec} represent programs as distributed vectors that capture syntactic and semantic information from the abstract syntax tree (AST) or intermediate representation (IR). The surge in deep learning has enabled feeding raw information such as AST \cite{alon2018general}, control-data flow graphs (CFG) \cite{brauckmann2020compiler, cummins2020programl}, and code token sequences \cite{cummins2017end} into powerful deep neural networks, capable of learning useful representations end-to-end.

\paragraph{Large Language Model} With the rise of large language models (LLMs), recent work has introduced LLMs into compiler optimization problems, demonstrating remarkable effectiveness in predicting optimal optimization sequences and comprehending the optimization process \cite{cummins2023large}. LLMs can directly analyze the intermediate representation of code in textual form, enabling a better understanding of the code's intricate details. However, due to limitations on input length, it has to split programs into shorter functions for training and prediction, preventing it from predicting on code of arbitrary length. We also want to emphasize that optimizing individual functions is quite different from optimizing an entire program. Many optimization passes, such as \textit{--inline} and \textit{--mergefunc}, which handle function inlining and merging, are either ineffective or less critical when there's only one function to optimize.

\paragraph{Frameworks} Platforms that expose the compiler as a playground for AI experiments have significantly reduced the entry barriers to intelligent compiler research. OpenTuner \cite{ansel2014opentuner}, and YaCoS \cite{filho2018yet} serve as autotuning frameworks with a range of compiler optimization techniques. Our experiments utilize CompilerGym \cite{cummins2022compilergym}, which offers user-friendly interfaces for researchers to interact with compilers in a reinforcement learning manner. We are optimistic that the future release of our compiler world model and code optimization agents, in conjunction with these platforms, can have a democratizing effect on applying AI techniques to compiler optimizations.

\section{Implementation Details}
\label{app:all_details}
\subsection{Compiler Environment}

Our experiments are conducted on the CompilerGym platform \cite{cummins2022compilergym}, version 0.2.5, with LLVM-10.0.0 integration.

\subsection{Details of Features and Actions}
\label{app:feature}
As shown in Table~\ref{tab:obs_action}, we adpot the 56-dimensional Autophase feature and a reduced action space from Autophase \cite{huang2019autophase}. A detailed list of the Autophase features and actions space is available in Table~\ref{tab:autophase-detail} and Table~\ref{tab:autophase-actions}.

\begin{table*}[tbp]
\centering
\caption{Descriptions of 56-dimension Autophase features (adapted from \cite{cummins2022compilergym}).}
\label{tab:autophase-detail}
\begin{tabular}{lll}
\toprule
{\textbf{Index}} & {\textbf{Name}}        & {\textbf{Description}}                                             \\
\midrule
{0}              & {BBNumArgsHi}          & {Number of BB where total args   for phi nodes is gt 5}            \\
{1}              & {BBNumArgsLo}          & {Number of BB where total args   for phi nodes is {[}1, 5{]}}      \\
{2}              & {onePred}              & {Number of basic blocks with 1   predecessor}                      \\
{3}              & {onePredOneSuc}        & {Number of basic blocks with 1   predecessor and 1 successor}      \\
{4}              & {onePredTwoSuc}        & {Number of basic blocks with 1   predecessor and 2 successors}     \\
{5}              & {oneSuccessor}         & {Number of basic blocks with 1   successor}                        \\
{6}              & {twoPred}              & {Number of basic blocks with 2   predecessors}                     \\
{7}              & {twoPredOneSuc}        & {Number of basic blocks with 2   predecessors and 1 successor}     \\
{8}              & {twoEach}              & {Number of basic blocks with 2   predecessors and successors}      \\
{9}              & {twoSuccessor}         & {Number of basic blocks with 2   successors}                       \\
{10}             & {morePreds}            & {Number of basic blocks with gt.   2 predecessors}                 \\
{11}             & {BB03Phi}              & {Number of basic blocks with Phi   node count in range (0, 3{]}}   \\
{12}             & {BBHiPhi}              & {Number of basic blocks with   more than 3 Phi nodes}              \\
{13}             & {BBNoPhi}              & {Number of basic blocks with no   Phi nodes}                       \\
{14}             & {BeginPhi}             & {Number of Phi-nodes at   beginning of BB}                         \\
{15}             & {BranchCount}          & {Number of branches}                                               \\
{16}             & {returnInt}            & {Number of calls that return an   int}                             \\
{17}             & {CriticalCount}        & {Number of critical edges}                                         \\
{18}             & {NumEdges}             & {Number of edges}                                                  \\
{19}             & {const32Bit}           & {Number of occurrences of 32-bit   integer constants}              \\
{20}             & {const64Bit}           & {Number of occurrences of 64-bit   integer constants}              \\
{21}             & {numConstZeroes}       & {Number of occurrences of   constant 0}                            \\
{22}             & {numConstOnes}         & {Number of occurrences of   constant 1}                            \\
{23}             & {UncondBranches}       & {Number of unconditional   branches}                               \\
{24}             & {binaryConstArg}       & {Binary operations with a   constant operand}                      \\
{25}             & {NumAShrInst}          & {Number of AShr instructions}                                      \\
{26}             & {NumAddInst}           & {Number of Add instructions}                                       \\
{27}             & {NumAllocaInst}        & {Number of Alloca instructions}                                    \\
{28}             & {NumAndInst}           & {Number of And instructions}                                       \\
{29}             & {BlockMid}             & {Number of basic blocks with   instructions between {[}15, 500{]}} \\
{30}             & {BlockLow}             & {Number of basic blocks with   less than 15 instructions}          \\
{31}             & {NumBitCastInst}       & {Number of BitCast instructions}                                   \\
{32}             & {NumBrInst}            & {Number of Br instructions}                                        \\
{33}             & {NumCallInst}          & {Number of Call instructions}                                      \\
{34}             & {NumGetElementPtrInst} & {Number of GetElementPtr   instructions}                           \\
{35}             & {NumICmpInst}          & {Number of ICmp instructions}                                      \\
{36}             & {NumLShrInst}          & {Number of LShr instructions}                                      \\
{37}             & {NumLoadInst}          & {Number of Load instructions}                                      \\
{38}             & {NumMulInst}           & {Number of Mul instructions}                                       \\
{39}             & {NumOrInst}            & {Number of Or instructions}                                        \\
{40}             & {NumPHIInst}           & {Number of PHI instructions}                                       \\
{41}             & {NumRetInst}           & {Number of Ret instructions}                                       \\
{42}             & {NumSExtInst}          & {Number of SExt instructions}                                      \\
{43}             & {NumSelectInst}        & {Number of Select instructions}                                    \\
{44}             & {NumShlInst}           & {Number of Shl instructions}                                       \\
{45}             & {NumStoreInst}         & {Number of Store instructions}                                     \\
{46}             & {NumSubInst}           & {Number of Sub instructions}                                       \\
{47}             & {NumTruncInst}         & {Number of Trunc instructions}                                     \\
{48}             & {NumXorInst}           & {Number of Xor instructions}                                       \\
{49}             & {NumZExtInst}          & {Number of ZExt instructions}                                      \\
{50}             & {TotalBlocks}          & {Number of basic blocks}                                           \\
{51}             & {TotalInsts}           & {Number of instructions (of all   types)}                          \\
{52}             & {TotalMemInst}         & {Number of memory instructions}                                    \\
{53}             & {TotalFuncs}           & {Number of non-external   functions}                               \\
{54}             & {ArgsPhi}              & {Total arguments to Phi nodes}                                     \\
{55}             & {testUnary}            & {Number of Unary operations}         \\                             
\bottomrule
\end{tabular}

\end{table*}

\begin{table*}[tbp]
\setlength{\tabcolsep}{1mm}
\centering
\caption{A list of LLVM transformation passes selected in Autophase's action space.}
\label{tab:autophase-actions}
\label{tab:passes}
\begin{tabular}{llllll}
\toprule
{\textbf{Index}} & {\textbf{Name}} & {\textbf{Index}} & {\textbf{Name}} & {\textbf{Index}} & {\textbf{Name}}  \\
\midrule
0 & -adce & 14 & -instcombine & 28 & -lowerinvoke \\
1 & -break-crit-edges & 15 & -ipsccp & 29 & -lowerswitch \\
2 & -constmerge & 16 & -jump-threading & 30 & -mem2reg \\
3 & -correlated-propagation & 17 & -lcssa & 31 & -memcpyopt \\
4 & -deadargelim & 18 & -licm & 32 & -partial-inliner \\
5 & -dse & 19 & -loop-deletion & 33 & -prune-eh \\    
6 & -early-cse & 20 & -loop-idiom & 34 & -reassociate \\
7 & -functionattrs & 21 & -loop-reduce & 35 & -sccp \\
8 & -functionattrs & 22 & -loop-rotate & 36 & -simplifycfg \\
9 & -globaldce & 23 & -loop-simplify & 37 & -sink \\
10 & -globalopt & 24 & -loop-unroll & 38 & -sroa \\
11 & -gvn & 25 & -loop-unswitch & 39 & -strip \\
12 & -indvars & 26 & -lower-expect & 40 & -strip-nondebug \\
13 & -inline & 27 & -loweratomic & 41 & -tailcallelim \\
\bottomrule
\end{tabular}

\end{table*}

\subsection{Hyperparameters}
\paragraph{CompilerDream}
The hyperparameters for our world model and agent implementation are outlined in Table~\ref{tab:hyperparameter}. For hyperparameters not specified, we use the same value as the DreamerV3 \cite{hafner2023mastering}. Most of the listed hyperparameter values are directly taken from DreamerV2 \cite{hafnermastering} or previous works, with minimal tuning. However, the \textit{loss scales} are carefully tuned, as the observation loss is relatively small compared to the reward loss in our setting, and the balance between these losses has a significant impact on CompilerDream's performance.
All experiments share the same set of hyperparameters unless otherwise specified.
\paragraph{Model-based Baseline}
For MBPO in Section~\ref{sec:analysis}, the dynamics model is a 4-layer MLP with a hidden size of 1024. We adopt the implementation from \href{https://github.com/Xingyu-Lin/mbpo_pytorch}{https://github.com/Xingyu-Lin/mbpo\_pytorch}, with most hyperparameters left unchanged. The modified hyperparameters are listed in Table~\ref{tab:mbpo-params}.

\begin{table}[tbp]

\centering
\caption{Hyperparameters in our experiments.}
\label{tab:hyperparameter}
\setlength{\tabcolsep}{3pt}
\footnotesize
\begin{tabular}{lll}
\toprule
                              & Hyperparameter          & \multicolumn{1}{l}{Value}           \\ \midrule
\multirow{6}{*}{\makecell[l]{Architecture}}         
                              & RSSM recurrent units                & 1024                                \\
                              & RSSM number of latents                & 32                                \\
                              & RSSM classes per latent                & 32                                \\
                              & MLP layers                & 4                                \\
                              & MLP hidden units                & 400                                \\
                              & Activation & LayerNorm + SiLU \\ \midrule
\multirow{21}{*}{\makecell[l]{Training}} 
                              
                              & Random exploration           & 500 environment steps \\
                              & Replay buffer capacity        & $2\times 10^6$    \\
                              & Reward smoothing $\alpha$ \cite{lee2023dreamsmooth} & 0.6 \\
                              & Training frequency           & Every 5 environment steps\\
                              & Batch size        & 50                           \\
                              & Batch length $T$   &   50                              \\
                              & Imagination horizon $H$ & 15 \\
                              & Discount    $\gamma$   & 0.99                                  \\
                              & $\lambda$-target discount & 0.95 \\
                              & World model loss scales & 100.0 for Autophase\\
                              & & 10.0 for action histogram \\
                              & & 1.0 for reward \\
                              & & 5.0 for discount \\
                              & & 0.1 for KL \\
                              & Actor entropy regularization $\eta$ & $3\times 10^{-4}$ \\
                              & KL balancing & 0.8 \\
                              & Optimizer                     & Adam                    \\
                             & World model learning rate     & $1\times 10^{-4}$       \\
                             & Actor-critic learning rate    & $3\times 10^{-5}$       \\
                             & Weight decay                  & $1\times 10^{-5}$       \\
                             & Gradient clipping & 100 \\ \bottomrule
\end{tabular}

\end{table}

\paragraph{Model-free Baselines}

We use RLlib \cite{liang2018rllib} to train and test model-free reinforcement learning algorithms including PPO \cite{schulman2017proximal} \cite{huang2019autophase}, A2C \cite{mnih2016asynchronous}, IMPALA \cite{espeholt2018impala}, APEX \cite{horgan2018distributed}, and DQN \cite{mnih2015human}. We use default hyperparameters of algorithms in RLlib following the CompilerGym platform \cite{cummins2022compilergym}, except that we have carefully tuned the hyperparameters for PPO, as listed in Table \ref{tab:rllib-params}. We explored roughly 2 or 3 values for each hyperparameter listed and selected the set of hyperparameters that yielded the best performance on the validation set. The results of the PPO baseline in Section \ref{sec:analysis} and Figure \ref{fig:ablation} are exactly the same as those in Section \ref{sec:main_exp} and Figure \ref{fig:results}.

\begin{table}[t]
\centering
\caption{Hyperparameters for the PPO baseline, well-tuned on our dataset to be deviating from the default value in RLlib.}
\label{tab:rllib-params}
\begin{tabular}{lll}
\toprule
                      & Hyperparameters      & Value  \\ 
\midrule
\multirow{13}{*}{PPO} & \texttt{gamma}                & 1.0    \\
                      & \texttt{use\_gae}             & True   \\
                      & \texttt{lambda\_}             & 1.0    \\
                      & \texttt{train\_batch\_size}   & 9000   \\
                      & \texttt{lr}                   & 5e-5 \\
                      & \texttt{kl\_coeff}            & 0.2    \\
                      & \texttt{kl\_target}           & 0.01   \\
                      & \texttt{vf\_loss\_coeff}      & 1.0    \\
                      & \texttt{num\_sgd\_iter}       & 30     \\
                      & \texttt{sgd\_minibatch\_size} & 128    \\
                      & \texttt{clip\_param}          & 0.3    \\
                      & \texttt{vf\_clip\_param}      & 10.0   \\
                      & \texttt{weight\_decay}        & 1e-6 \\
\bottomrule                      
\end{tabular}

\end{table}

\begin{table}[t]
\centering
\caption{Modified hyperparameters in our MBPO baseline.}
\label{tab:mbpo-params}
\setlength{\tabcolsep}{3pt}
\begin{tabular}{lll}
\toprule
                      & Hyperparameters      & Value  \\ 
\midrule
\multirow{5}{*}{MBPO}     & \texttt{target\_update\_interval}   & 10 \\
                         & \texttt{lr}      & 1e-3 \\
                          & \texttt{rollout\_batch\_size}       & 50000 \\
                       & \texttt{rollout\_max\_length} & 45\\
                          & \texttt{init\_exploration\_steps}      & 3600 \\
\bottomrule                      
\end{tabular}

\end{table}


\subsection{Hardware and Training Time}
\paragraph{CompilerDream}
We train all CompilerDream-based methods with 64 CPUs and an RTX-3090 GPU. In Section \ref{sec:main_exp}, we trained CompilerDream on the CodeContests dataset for around 1 day and 20 hours. In Sections \ref{sec:search_exp} and \ref{sec:coreset_exp}, CompilerDream was trained for about 1 day.
\paragraph{Random Search}
The random search baseline in Figure \ref{fig:results} is conducted with 4 CPUs, which is sufficient since it is a single-thread program. 
The random search baseline in Table \ref{tab:leaderboard} utilizes 80 CPUs, as specified in the write-up attached to CompilerGym's leaderboard \cite{compilergymleaderboard}. Our \textit{CompilerDream + Guided Search} method in Table \ref{tab:leaderboard} is tested on the same machine used for training CompilerDream, equipped with 64 CPUs and an RTX-3090 GPU.
\paragraph{Model-free and Model-based baselines}

The hardware setting is the same as used for training CompilerDream. For the PPO baseline in Section~\ref{sec:main_exp}, we train for about 7 hours, as longer training leads to overfitting on CodeContests. Other model-free methods in Section~\ref{sec:main_exp} are trained for at least 10 hours. We use 5 workers for environment interaction and 4 evaluation workers to assess checkpoints on the validation set. In Section~\ref{sec:analysis}, we train MBPO for approximately 10 hours, as its evaluation score converges within this time.

\subsection{Random Seeds}
All experiments reporting a min-max range or standard deviation are conducted with three different random seeds. The results in Table~\ref{tab:coreset}, Table~\ref{tab:compare_mbpo}, and Table~\ref{tab:reward_smoothing} are also averaged over three runs with different seeds. The seeds are randomly selected, and the results are generally consistent across other random seeds as well.

\subsection{Detail of Guided Search in Autotuning}
\label{app:search}
Our guided search in Section~\ref{sec:search_exp} follows the \textit{PPO+Guided Search} design. Actions are sampled from the learned policy for up to 45 steps per episode, recording code size reductions after each step to track the maximum. To encourage exploration, 5\% of actions are sampled uniformly at random. We record the best reduction so far and monitor elapsed time to stay within the 1-minute budget per benchmark. Unlike \textit{PPO+Guided Search} (Table~\ref{tab:leaderboard}), which uses a 200-step horizon and an extra 500-step search on the best sequence, we omit the latter and use only 45 steps, making our wall time slightly shorter.

\subsection{Detail of Value Prediction Experiment}
\label{app:coreset}

We provide additional details for the experiment in Section~\ref{sec:coreset_exp}. The baseline, Coreset-NVP\cite{liang2023rlcompopt}, trains a value model to score sequences from a fixed \textit{core set}, obtained via extensive search on its training set. For each program, it evaluates the top-scored sequences in order, executing up to 45 passes across them. After each full sequence, the IR resets to the initial state. Code size reductions after each pass are tracked, and the best-performing pass and prefix are selected. The core set includes 50 sequences (625 passes total, 12.5 passes each on average), so around 3–4 sequences are tried per benchmark. The 79 passes used correspond to the optimization actions in Section~\ref{sec:pomdp}.

In our setup, the action space includes all 124 LLVM passes. To train the world model to predict cumulative reward values for the core set passes, we apply each core set sequence to training programs and collect trajectories. We use the CodeContests dataset (Section~\ref{sec:train_dataset}), consistent with the experiments in Section~\ref{sec:main_exp}. Reward loss weight is set to 100.0, with other hyperparameters unchanged. For evaluation, we compile the top 45 predicted prefixes per program and report the best.

\subsection{AI-Generated Benchmarks}
\label{app:compilergym_benchmarks}

To further test the generalization ability of our CompilerDream agent on different programming languages, we borrow the method from FormAI \cite{tihanyi2023formai} and generate a dataset containing 50 unique Objective-C programs using GPT-3.5. We use the same prompt as FormAI, except that we add an instruction to ask GPT to generate programs that can be directly compiled under Clang version 10.0.0 and do not use ARC (Automatic Reference Counting), to improve the compilation pass rate of generated programs. We compile the generated programs using Clang without including any third-party libraries, and all programs that cannot pass compilation are discarded. 

\section{Extended Experimental Results}

\begin{table*}[tbp]

\centering
\caption{Top performances of the zero-shot CompilerDream RL agent on individual programs. }
\label{tab:showcase}
\setlength{\tabcolsep}{1mm}
\begin{tabular}{cclccc}
\toprule
\multirow{2}{*}{Dataset}    & \multicolumn{1}{c}{\multirow{2}{*}{Program}} & \multicolumn{1}{c}{\multirow{2}{*}{Pass Sequences by CompilerDream Agent}} & \multicolumn{3}{c}{Code size}      \\
                            & \multicolumn{1}{c}{}  & \multicolumn{1}{c}{}  & \multicolumn{1}{c}{\texttt{O0}} & \multicolumn{1}{c}{CompilerDream} & \multicolumn{1}{c}{\texttt{Oz} (Reduction)} \\ \midrule
\multirow{12}{*}{cBench}    & sha  &  \makecell[l]{-sroa -gvn -sroa -sroa -instcombine -simplifycfg\\ -sroa -functionattrs -licm  -functionattrs -early-cse\\ -simplifycfg  -indvars -gvn -simplifycfg -early-cse \\-functionattrs -reassociate -memcpyopt -gvn\\ -simplifycfg -reassociate -functionattrs -early-cse \\-instcombine -early-cse -lowerinvoke -simplifycfg\\ -functionattrs -functionattrs -memcpyopt -gvn \\ -simplifycfg -functionattrs -reassociate -early-cse  \\ -instcombine -early-cse -strip -strip \\-early-cse -strip -early-cse -strip -early-cse}                                                        &   799  &    349     &   500 (\textbf{1.43$\times$})   \\ \cmidrule{2-6} 
                            &  bzip2  & \makecell[l]{
                            -sroa -gvn -sroa -sroa -simplifycfg -strip -strip\\ 
                            -instcombine -simplifycfg -early-cse -simplifycfg \\
                            -strip -strip -reassociate -reassociate -memcpyopt \\
                            -jump-threading -memcpyopt -functionattrs \\
                            -memcpyopt -functionattrs -licm -functionattrs -gvn \\
                            -simplifycfg -functionattrs -functionattrs -reassociate \\
                            -memcpyopt -early-cse -jump-threading -memcpyopt \\
                            -early-cse -strip -strip -instcombine -simplifycfg \\
                            -jump-threading -jump-threading -early-cse \\
                            -jump-threading -early-cse -simplifycfg 
                            -strip \\ -functionattrs}  &   28748     &   13565   &  15946 (\textbf{1.18$\times$})   \\  
                            \midrule
\multirow{12}{*}{OpenCV}     &   \#41    &    \makecell[l]{
                            -sroa -early-cse -early-cse -simplifycfg -early-cse\\
                            -early-cse -early-cse -sink -sink -sink -loop-deletion \\
                            -loop-deletion -loop-deletion -lowerinvoke -simplifycfg\\
                            -sroa -sroa -sroa -early-cse -early-cse -early-cse\\
                            -early-cse -lowerinvoke -lowerinvoke -lowerinvoke \\
                            -globalopt -sroa -instcombine -simplifycfg -early-cse\\
                            -early-cse -early-cse -sink -globalopt -sroa -sroa \\
                            -functionattrs -simplifycfg -early-cse -functionattrs\\
                            -early-cse -sink -functionattrs -functionattrs\\
                            -functionattrs
                            }  &    28    &   18      &   28 (\textbf{1.56$\times$})     
                            \\ \cmidrule{2-6} 
                            &   \#9    &   \makecell[l]{
                            -sroa -gvn -sroa -early-cse -simplifycfg -strip -strip\\
                            -loop-deletion -lowerinvoke -simplifycfg -sroa \\
                            -early-cse -instcombine -simplifycfg -early-cse \\
                            -memcpyopt -early-cse -strip -strip -strip -functionattrs\\
                            -instcombine -simplifycfg -jump-threading -simplifycfg\\
                            -memcpyopt -early-cse -simplifycfg -reassociate\\
                            -memcpyopt -functionattrs -functionattrs -reassociate\\
                            -memcpyopt -early-cse -instcombine -simplifycfg \\
                            -jump-threading -early-cse -jump-threading -strip \\
                            -simplifycfg -early-cse -simplifycfg -functionattrs}    &  9510    &   6341    &   9269 (\textbf{1.18$\times$}) \\ \midrule
\multirow{2}{*}{TensorFlow} &  \#17   &   \makecell[l]{
                            -sroa -gvn -early-cse -simplifycfg -early-cse -early-cse \\
                            -instcombine -early-cse -sink -sink -sink -lowerinvoke\\
                            -simplifycfg -sroa -early-cse -memcpyopt -early-cse \\
                            -memcpyopt -instcombine -simplifycfg -early-cse \\
                            -early-cse -early-cse -simplifycfg -early-cse -early-cse\\
                            -strip -strip -indvars -simplifycfg -early-cse -memcpyopt\\
                            -early-cse -instcombine -early-cse -early-cse -simplifycfg\\
                            -early-cse -memcpyopt -early-cse -early-cse -instcombine\\ 
                            -simplifycfg -early-cse -early-cse}   & 5512  &    4247   &   5450 (\textbf{1.28$\times$})  
                            \\ \bottomrule
\end{tabular}

\end{table*} 

\subsection{Learning Curves}

\begin{figure}[htb]
    \centering
    \includegraphics[width=\linewidth]{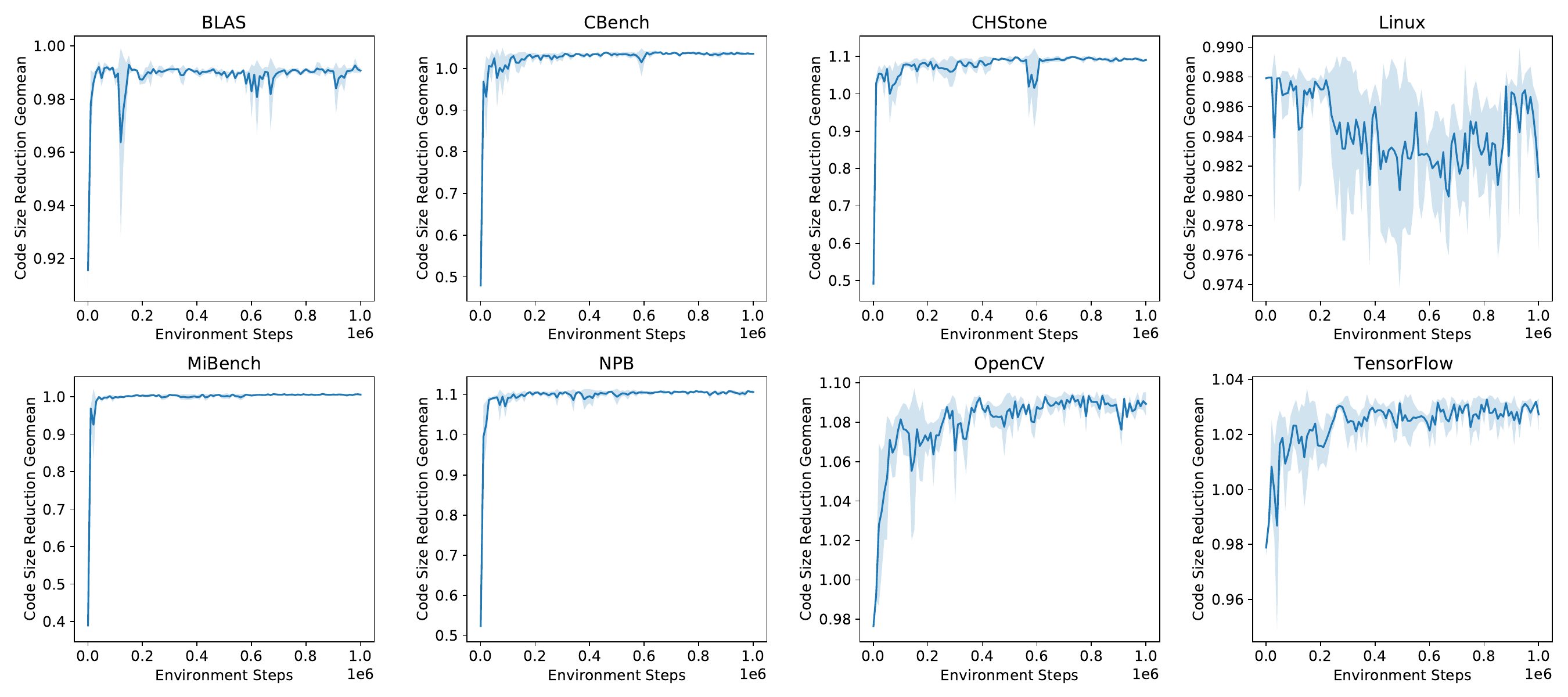}
    \caption{Zero-shot test performance of the CompilerDream agent during training. We report mean and standard deviation across three runs.}
    \Description{The learning curves for 7 out of the 8 datasets saturate quickly, while performance on the Linux dataset remains almost constant throughout the training process.}
    \label{fig:curves}
\end{figure}

The performance of our CompilerDream with RL agents during training is shown in Figure~\ref{fig:curves}. Note that we validate and test the agent every 10000 environment steps and report the test performance from the checkpoint that achieved the best validation results for comparison among various methods. 

\begin{table*}[tbh]
\centering
\caption{Quantitative results for code size reduction, corresponding to Figure \ref{fig:results}. We report mean and standard deviation across three runs.}
\label{tab:quantitative}
\setlength{\tabcolsep}{1mm}
\begin{tabular}{llcccccc}
\toprule
 &  & -O0 & Random Policy & \multicolumn{1}{c}{\begin{tabular}[c]{@{}c@{}}Random Search\\ (1s)\end{tabular}} & Autophase & \multicolumn{1}{c}{\begin{tabular}[c]{@{}c@{}}CompilerDream\\ (In-domain)\end{tabular}} & \multicolumn{1}{c}{\begin{tabular}[c]{@{}c@{}}CompilerDream\\ (Zero-shot)\end{tabular}}\\
\midrule
\multirow{3}{*}{BLAS} 
 & geomean & 0.931 & 0.821$\pm$0.018 & 0.960$\pm$0.004 & \textbf{0.993}$\pm$0.002 & \textbf{0.993}$\pm$0.005 & 0.991$\pm$0.000\\
 & min & 0.707 & 0.432$\pm$0.106 & 0.723$\pm$0.076 & 0.916$\pm$0.004 & 0.831$\pm$0.148 & 0.913$\pm$0.000\\
 & max & 1.000 & 1.002$\pm$0.004 & 1.031$\pm$0.010 & 1.014$\pm$0.001 & 1.025$\pm$0.005 & 1.016$\pm$0.000
\\
\midrule
\multirow{3}{*}{cBench} 
 & geomean & 0.481 & 0.737$\pm$0.015 & 0.858$\pm$0.020 & 1.010$\pm$0.004 & N/A & \textbf{1.036}$\pm$0.002\\
 & min & 0.299 & 0.401$\pm$0.097 & 0.560$\pm$0.030 & 0.817$\pm$0.109 & N/A & 0.962$\pm$0.001\\
 & max & 0.626 & 1.150$\pm$0.113 & 1.298$\pm$0.083 & 1.405$\pm$0.014 & N/A & 1.395$\pm$0.054
\\
\midrule
\multirow{3}{*}{CHStone} 
 & geomean & 0.487 & 0.701$\pm$0.043 & 1.037$\pm$0.006 & 1.071$\pm$0.002 & N/A & \textbf{1.094}$\pm$0.005\\
 & min & 0.402 & 0.278$\pm$0.108 & 0.889$\pm$0.044 & 0.967$\pm$0.027 & N/A & 1.000$\pm$0.006\\
 & max & 0.655 & 1.038$\pm$0.033 & 1.304$\pm$0.038 & 1.313$\pm$0.045 & N/A & 1.378$\pm$0.021
\\
\midrule
\multirow{3}{*}{Linux} 
 & geomean & 0.988 & 0.981$\pm$0.004 & 1.001$\pm$0.000 & 0.985$\pm$0.001 & \textbf{0.993}$\pm$0.000 & 0.986$\pm$0.005\\
 & min & 0.615 & 0.740$\pm$0.126 & 1.000$\pm$0.000 & 0.615$\pm$0.000 & 0.700$\pm$0.001 & 0.642$\pm$0.041\\
 & max & 1.011 & 1.001$\pm$0.002 & 1.011$\pm$0.000 & 1.009$\pm$0.003 & 1.011$\pm$0.000 & 1.009$\pm$0.003
\\
\midrule
\multirow{3}{*}{Mibench} 
 & geomean & 0.389 & 0.812$\pm$0.022 & 1.005$\pm$0.001 & 1.000$\pm$0.007 & N/A & \textbf{1.006}$\pm$0.002\\
 & min & 0.278 & 0.352$\pm$0.074 & 0.858$\pm$0.010 & 0.833$\pm$0.062 & N/A & 0.879$\pm$0.003\\
 & max & 0.760 & 1.392$\pm$0.074 & 1.603$\pm$0.015 & 1.524$\pm$0.045 & N/A & 1.588$\pm$0.000
\\
\midrule
\multirow{3}{*}{NPB} 
 & geomean & 0.530 & 0.833$\pm$0.011 & 1.074$\pm$0.010 & 1.071$\pm$0.018 & 1.075$\pm$0.017 & \textbf{1.108}$\pm$0.001\\
 & min & 0.191 & 0.414$\pm$0.043 & 0.805$\pm$0.104 & 0.801$\pm$0.070 & 0.810$\pm$0.082 & 0.886$\pm$0.000\\
 & max & 1.066 & 1.848$\pm$0.193 & 2.141$\pm$0.062 & 2.235$\pm$0.101 & 2.315$\pm$0.028 & 2.343$\pm$0.012
\\
\midrule
\multirow{3}{*}{OpenCV} 
 & geomean & 0.981 & 0.949$\pm$0.011 & 1.080$\pm$0.001 & 1.082$\pm$0.004 & 1.087$\pm$0.007 & \textbf{1.092}$\pm$0.000\\
 & min & 0.833 & 0.603$\pm$0.133 & 0.888$\pm$0.021 & 0.897$\pm$0.002 & 0.898$\pm$0.001 & 0.897$\pm$0.000\\
 & max & 1.370 & 1.409$\pm$0.167 & 1.571$\pm$0.022 & 1.620$\pm$0.193 & 1.635$\pm$0.186 & 1.556$\pm$0.000
\\
\midrule
\multirow{3}{*}{TensorFlow} 
 & geomean & 0.983 & 0.912$\pm$0.009 & 1.006$\pm$0.003 & 1.010$\pm$0.019 & \textbf{1.032}$\pm$0.000 & \textbf{1.032}$\pm$0.001\\
 & min & 0.927 & 0.618$\pm$0.046 & 0.877$\pm$0.009 & 0.841$\pm0$.134 & 0.970$\pm$0.010 & 0.968$\pm$0.011\\
 & max & 1.010 & 1.235$\pm$0.034 & 1.267$\pm$0.007 & 1.281$\pm$0.006 & 1.289$\pm$0.008 & 1.282$\pm$0.002
\\
\bottomrule
\end{tabular}

\end{table*}

\subsection{Quantitative Results}

Quantitative results corresponding to Figure 3 in the main text are provided in Table~\ref{tab:quantitative}.

\subsection{Program Case Study}
In Addition to the predicted optimization trajectory showcased in our paper, we present more examples of optimization results here. Table \ref{tab:showcase} presents the RL agent's top performance outcomes on various benchmark datasets. We observe that the agent indeed produces a specialized optimization strategy tailored for each program. Additionally, these results also highlight certain passes, such as \textit{-sroa} (scalar replacement of aggregates), \textit{-early-cse}(early common subexpression elimination), and \textit{-simplifycfg}, as particularly effective in code size optimization.

\begin{figure}[htb]
    \centering
    \includegraphics[width=1.0\linewidth]{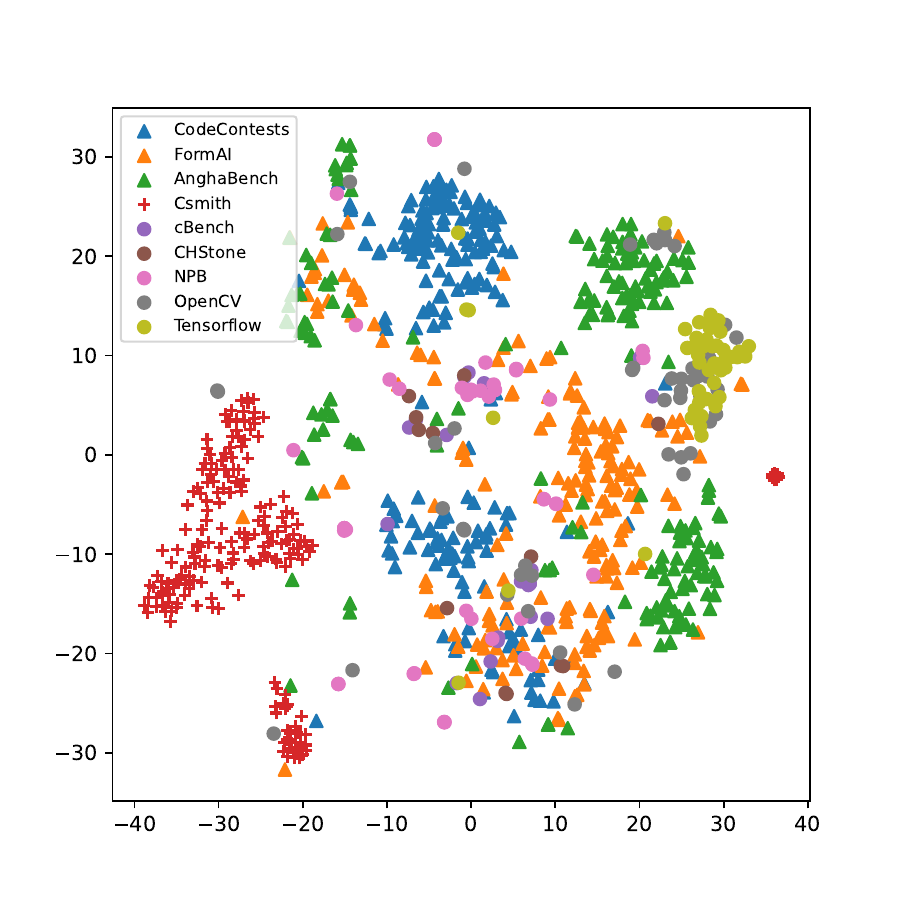}
    \caption{t-SNE \cite{van2008visualizing} visualization of programs from training and test datasets.}
    \label{fig:tsne}
    \Description{In each dataset, the majority of data points are concentrated within a few clusters, with a smaller proportion dispersed across various regions. Notably, the data distribution in the CodeContest dataset overlaps with that of most other datasets.}
\end{figure} 
\subsection{Data Distribution Visualization}

In Figure~\ref{fig:tsne}, we visualize the distribution of our training and test datasets. 
To accurately represent the dynamic behavior of programs, we randomly select 1000 action sequences, each with a length of 45, from our action space. These sequences are subsequently executed on each program, with the resulting Autophase features concatenated to form a feature vector with dimensions of $1000\times 45\times 56$ for every program. These comprehensive feature vectors are finally processed using t-SNE \cite{van2008visualizing} for dimensionality reduction and visualization.

Figure~\ref{fig:tsne} illustrates that our training data (denoted as triangles) has a broad coverage of test programs (denoted as circles). This contrasts with the Csmith dataset (denoted as crosses) employed in CompilerGym experiments \cite{cummins2022compilergym}, which shows a significant deviation from real-world applications. Nonetheless, our visualization can still not perfectly capture the transferability across datasets. For instance, empirical evidence suggests that CodeContests are the most effective in generalizing to OpenCV and TensorFlow, while the visual analysis does not directly imply this.

\end{document}